\documentclass[fleqn,usenatbib]{mnras}

\usepackage{newtxtext,newtxmath}
\usepackage{array}
\usepackage{tabularx}

\usepackage[T1]{fontenc}

\DeclareRobustCommand{\VAN}[3]{#2}
\let\VANthebibliography\thebibliography
\def\thebibliography{\DeclareRobustCommand{\VAN}[3]{##3}\VANthebibliography}

\usepackage{graphicx}	
\usepackage{amsmath}	

\title[DESI EDR photo-$z$.]{Estimating Photometric Redshifts for Galaxies from the DESI Legacy Imaging Surveys with Bayesian Neural Networks Trained by DESI EDR}

\author[X. Zhou et al.]{
Xingchen Zhou$^{1,2}$,
Nan Li$^{1, 2}$\thanks{E-mail: nan.li@nao.cas.cn},
Hu Zou$^{1}$,
Yan Gong$^{1, 3, 2}$,
Furen Deng$^{1, 3}$,
Xuelei Chen$^{1}$,
Qian Yu$^{1}$, 
Zizhao He$^{4}$, \newauthor
Boyi Ding$^{5}$
\\
$^{1}$National Astronomical Observatories, CAS, Datun road, 20, Chaoyang, Beijing 100101, China\\
$^{2}$Science Center for China Space Station Telescope, National Astronomical Observatories, Chinese Academy of Sciences, 20A Datun Road, Beijing 100101, China\\
$^{3}$University of Chinese Academy of Sciences, Beijing 100049, China\\
$^{4}$Purple Mountain Observatory, Chinese Academy of Sciences, Nanjing, Jiangsu, 210023, China\\
$^{5}$Leiden Observatory, Leiden University, P.O. Box 9513, 2300 RA Leiden, The Netherlands
}

\date{Accepted XXX. Received YYY; in original form ZZZ}

\pubyear{2015}

\begin{document}
\label{firstpage}
\pagerange{\pageref{firstpage}--\pageref{lastpage}}
\maketitle

\begin{abstract}
We present a catalogue of photometric redshifts for galaxies from DESI Legacy Imaging Surveys, which includes $\sim0.18$ billion sources covering 14,000 $\deg^2$. The photometric redshifts, along with their uncertainties, are estimated through galaxy images in three optical bands ($g$, $r$ and $z$) from DESI and two near-infrared bands ($W1$ and $W2$) from WISE using a Bayesian Neural Network (BNN). The training of BNN is performed by above images and their corresponding spectroscopic redshifts given in DESI Early Data Release (EDR). Our results show that categorizing galaxies into individual groups based on their inherent characteristics and estimating their photo-$z$s within their group separately can effectively improve the performance. Specifically, the galaxies are categorized into four distinct groups based on DESI's target selection criteria: Bright Galaxy Sample (BGS), Luminous Red Galaxies (LRG), Emission Line Galaxies (ELG) and a group comprising the remaining sources, referred to as NON. {As measured by outliers of $|\Delta z| > 0.15 (1 + z_{\rm true})$, accuracy $\sigma_{\rm NMAD}$ and mean uncertainty $\overline{E}$ for BNN, we achieve low outlier percentage, high accuracy and low uncertainty: 0.14\%, 0.018 and 0.0212 for BGS and 0.45\%, 0.026 and 0.0293 for LRG respectively, surpassing results without categorization.} {However, the photo-$z$s for ELG cannot be reliably estimated, showing result of $>15\%$, $\sim0.1$ and $\sim0.1$ irrespective of training strategy.} {On the other hand, NON sources can reach $1.9\%$, 0.039 and 0.0445 when a magnitude cut of $z<21.3$ is applied.} Our findings demonstrate that estimating photo-$z$s directly from galaxy images is significantly potential, and to achieve high-quality photo-$z$ measurement for ongoing and future large-scale imaging survey, it is sensible to implement categorization of sources based on their characteristics. 
\end{abstract}


\begin{keywords}
methods: statistical – techniques: image processing – galaxies: distances and redshifts – galaxies: photometry
\end{keywords}



\section{Introduction}\label{sec: intro}
Redshift is a fundamental quantity in cosmological studies based on galaxy surveys. The most accurate redshift measurements are obtained by observing spectra for galaxies. However, this process is time-consuming and can no longer meet the demands for accurately measuring redshifts for millions of sources observed in current photometric surveys, let alone for next stage, more powerful and deeper ones like Euclid~\citep{Laureijs2011}, LSST~\citep{LSSTcoll2012} and CSST~\citep{Zhan2018}. Under such circumstances, photometric redshifts have become inevitable for most cosmological studies. While their accuracy may not match that of spectroscopic redshifts, their estimation speed is a significant advantage. Photometric redshifts rely on multi-band photometry, which captures the spectral energy distribution of galaxies without requiring individual spectra. These estimates are essential for large-scale surveys, where obtaining spectra for every source is impractical, for more detailed discussions, please refer to~\citet{Salvato2019, Newman2022, Brescia2021}.

Researchers are actively exploring methods to enhance the accuracy of photometric redshifts. The approaches to estimate photo-$z$s can mainly be divided into two categories. The first category is the fitting method, where spectral energy distribution (SED) templates  are utilized to fit the photometric measurements and determine the redshift that minimizes the $\chi^2$ value~\citep{Lanzetta1996,Fernandez1999,Arnouts1999,Bolzonella2000, Ilbert2006, Brammer2008}. This method is straightforward, however, the completeness of templates can significantly impact the performance, since low redshift and high redshift galaxies probably do not fit in same template sets. Another category is based on empirical methods, where relations between redshift and photometric measurements are established based on existing data with accurate redshift values. Machine learning (ML), particularly deep learning (DL) methods (also known as neural networks), is well-suited for implementing this approach~\citep{Firth2003, Tagliaferri2003, Collister2004, Sadeh2016}.
The multi-dimensional mapping from photometric measurements to redshifts is optimized by a large number of parameters in ML/DL model, resulting in improvement on accuracy compared to SED fitting provided that the galaxies and training ones span similar parameter spaces~\citep{Brescia2021}. Notably, deep learning has gained prominence due to the success of convolutional neural networks (CNNs) in computer vision tasks, outperforming traditional methods~\citep{Lecun1998, Krizhevsky2012}. Consequently, CNNs can be designed to directly predict photometric redshifts from multi-band imaging data, leveraging the advantages of not requiring explicit photometric measurements and naturally integrating morphological information from galaxy images~\citep{Pasquet2019, Henghes2022, Zhou2022ext, Zhou2022pho, Jones2023, Ait2024}. Particularly, research by~\citet{Zhou2022ext} demonstrates that galaxy images indeed offer additional information beyond photometric measurements, leading to a reduction in outlier percentage for photo-$z$ estimation. Unlike SED fitting, deep learning methods typically provide point estimates for photometric redshifts without uncertainties for each source. Recognizing the importance of uncertainty in cosmological studies, ~\citet{Zhou2022pho} and~\citet{Jones2023} are dedicated to provide both photometric redshift and uncertainty estimations employing Bayesian neural networks (BNN). Instead of having fixed values for the weights and biases of Bayesian network, these parameters are assigned with probability distributions, capturing their individual uncertainty. By propagating uncertainty through the network, not only point predictions but also confidence intervals or posterior distributions can be obtained~\citep{MacKay95:network, 2015arXiv150505424B, 2015arXiv150602142G}. 

Spectroscopic redshifts are required for both SED fitting and empirical methods, serving purposes as calibrating photometric measurements and training the model respectively. Acquiring sufficient sources with secure spectroscopic redshifts that covers similar parameter space to photometric sources is a challenging task. Fortunately, several ongoing and planned spectroscopic galaxy surveys, e.g. Dark Energy Spectroscopic Instrument (DESI)~\citep{DESIcoll2016}, Prime Focus Spectrograph (PFS)~\citep{Takada2014}, MUltiplexed Spectroscopic Telescope (MUST)\footnote{\url{https://must.astro.tsinghua.edu.cn/en}}, MegaMapper~\citep{Schlegel2022} and Wide-field Spectroscopic Telescope (WST)~\citep{Mainineri2024}, are aiming to provide a substantial amount of galaxy spectra with accurate spectroscopic redshifts. These datasets can be leveraged for calibration and training in both photometric redshift estimation methods.

The DESI project emerges as a notably ambitious endeavour in the field of spectroscopic experiments. Over its five-year mission, DESI aims to gather spectra for approximately 30 million galaxies, effectively covering one-third of the night sky. This comprehensive survey will transverse over 11 billion years of cosmic history, leveraging DESI's capabilities to study our universe through mechanisms such as baryon acoustic oscillations (BAO) and redshift-space distortions (RSD)~\citep{DESIcoll2016}. DESI will focus its observational efforts on four primary target categories: bright galaxy sample (BGS, $z < 0.6$), luminous red galaxies (LRG, $0.4 < z \sim 1.0$), emission line galaxies (ELG, $0.6 < z < 1.6$) and quasars (QSO, $z > 0.9$). These targets, which trace the dark matter distribution at increasing redshifts, are selected based on their optical characteristics in the $g$, $r$ and $z$ bands from the DESI Legacy Imaging Surveys (DESI LS; \citet{Dey2019}), and the near-infrared $W1$ and $W2$ bands from the Wide-field Infrared Survey Explorer (WISE; \citet{Wright2010}). The detailed strategies for preliminary and main target selection are extensively documented in \citet{Ruiz-macias2020,ZhouRP2020,Raichoor2020,Yeche2020} and \citet{Hahn2023,ZhouRP2023, Raichoor2023, Chaussidon2023}. Furthermore, DESI's Early Data Release (DESI EDR; \citet{2023arXiv230606308D})~\footnote{\url{https://data.desi.lbl.gov/doc/}} has already made available data on 1.2 million galaxies and quasars from Survey Validation (SV) observations. Many of these sources have secure spectroscopic redshifts, providing valuable dataset for training deep learning models aimed at estimating photometric redshifts. 

The DESI Legecy Imaging Surveys (DESI LS), as the foundation for target selection, covering approximately $14,000$ square degrees of sky, integrates data from three significant surveys: the Beijing-Arizona Sky Survey (BASS; ~\cite{2017PASP..129f4101Z, Zou2018}), the Mayall z-band Legacy Survey (MzLS; ~\citet{2016AAS...22831702S}), and the Dark Energy Camera Legacy Survey (DECaLS; ~\citet{2016AAS...22831701B}). BASS maps approximately $5,400$ square degrees of the northern sky in the $g$ and $r$ bands using the $2.3$ m Bok telescope at Kitt Peak. MzLS covers the same regions with its $z$ band observations on the $4$ m Mayall telescope at the same location. DECaLS, on the other hand, spans $9,000$ square degrees across both northern and southern skies, utilizing $g$, $r$ and $z$ bands of the $4$ m Blanco telescope at CTIO. Observations conducted with different instruments necessitate varying target selection criteria for the northern and southern skies. The all-sky survey by WISE in four near-infrared bands, with central wavelength as $3.4$, $4.6$, $12$, and $24\ \mu m$ for $W1$, $W2$, $W3$ and $W4$ respectively, is crucial for selecting targets among luminous red galaxies and quasars, whose photometric signatures are primarily determined by infrared observations~\citep{ZhouRP2020, Yeche2020}.

In this paper, we present a catalogue of photometric redshifts for galaxies from DESI Legacy Imaging Surveys. We use Bayesian neural network (BNN) to estimate photometric redshifts along with their uncertainties directly through galaxy images in three optical bands ($g$, $r$ and $z$) from DESI and two near-infrared bands ($W1$ and $W2$) from WISE. Notably, estimation from images has the advantage of naturally incorporating the morphological information. {Two Bayesian neural network configurations, MNF~\citep{2017arXiv170301961L} and MC-dropout~\citep{2015arXiv150602142G}, are investigated, and ultimately, we select the MNF models.
These models are trained using sources with secure spectroscopic redshifts from DESI EDR data.} Unlike previous photo-$z$ estimation efforts, sources are categorized into distinct groups based on their inherent charateristics and estimate their photo-$z$s within their corresponding group. This strategy can enhance the accuracy by effectively reducing potential confusion of sources in feature space. {Specifically, the sources are categorized into four groups, BGS, LRG, ELG and a group comprising the remaining sources, refered to as NON, based on DESI's main target selection criteria. Subsequently, their redshifts are estimated separately, resulting in enhanced performance compared to results without categorization, especially for BGS and LRG groups.} However, the performance of ELG and NON is not comparable to other groups. Therefore, we employ unsupervised clustering technique to investigate deeper causes of distinct performance for these four groups in feature space, and provide some guidance for improvements for NON sources. Additionally, the correlations between photo-$z$ precision and morphological classifications are also explored. Finally, we produce photometric redshift catalogue for BGS, LRG and part of NON sources considering their individual performance. The photometric redshift catalogue in this paper are published online\footnote{\url{https://pan.cstcloud.cn/s/hUWwk1QTSjo}}.

The structure of the paper is organized as follows: Section~\ref{sec: data} outlines the galaxy imaging and spectroscopic redshift data utilized in our work and the motivation for source categorization. Section~\ref{sec: methods} introduces neural network methods employed. The results for BGS, LRG, ELG and NON sources are presented in Section~\ref{sec: results}, where they are compared with previous studies. And in Section~\ref{sec: discussion}, we make some discussions, including some analysis with unsupervised clustering technique, the correlation of performance with morphological characteristics, and potential improvements with future data releases. This work is summarized in Section~\ref{sec: conclusion}. {Appendix~\ref{app: dropout} displays the results by BNNs on MC-dropout framework. Appendix~\ref{app: color-z} shows the correlations between the colors and spectroscopic redshifts, demonstrating the distinct behaviors for four group of sources. And Appendix~\ref{app: catalogue} describes our photo-$z$ catalogue.}

\section{Datasets} \label{sec: data}
In this section, we firstly describe the extraction of multi-band photometric imaging data from DESI LS and spectroscopic data from DESI EDR, and then explain the motivation that we categorize these sources into BGS, LRG, ELG and a group comprising the remaining sources (referred to as NON) based on DESI's target selection and estimate their photo-$z$ within their groups individually. 

\subsection{Photometric imaging data} \label{sec: image data}
We estimate photometric redshifts through galaxy images in three optical bands $g$, $r$ and $z$ from DESI and two near-infrared bands $W1$ and $W2$ bands from WISE. For each galaxy, three optical and two near-infrared images are both inputs for our deep learning model. Some diagnostic photometric features such as break or bump of galaxies at high redshift will leave their footprint at near-infrared bands, hence, the inclusion of these bands are crucial for precise photometric redshift estimations~\citep{LiuDZ2023}. 

The galaxy data are extracted from the DESI DR9 sweep catalogue~\footnote{\url{https://www.legacysurvey.org/dr9/files/\#sweep-catalogs-region-sweep}}, which contains a subset of the most commonly used photometric measurements by Tractor~\footnote{\url{https://github.com/dstndstn/tractor}} software. The morphological classification is also performed by this software to separate stars from galaxies, hence we only include sources that are not morphologically classified as point spread function (PSF). Additionally, galaxies lacking reliable optical and near-infrared observations or those within masked areas are excluded from our study. Finally, the constraints applied to the photometric data can be summarized as follows:
\begin{equation} \label{eq: photo selection}
\begin{split}
    &\qquad\qquad \rm TYPE\ != PSF\\
    &\qquad\qquad \rm FLUX\_G, R, Z, W1, W2 > 0\\
    &\qquad\qquad \rm FLUX\_IVAR\_G, R, Z, W1, W2 > 0\\
    &\qquad\qquad \rm MASKBITS == 0
\end{split}
\end{equation}

To facilitate the extraction and processing of galaxy images, we employ the Cutout2D class from the Astropy package~\footnote{\url{https://www.astropy.org/}}. Each image, with five bands, is configured to a standard size of $10\arcsec$ with the galaxy positioned at the center. We also investigate various sizes, like $10$, $20$ and $30\arcsec$, and found the $10\arcsec$ is the most optimal by achieving the best photo-$z$ estimation. While we admit that this size threshold may exclude some edge features in larger galaxies, potentially affecting the performance, the majority of galaxies in our study are smaller-sized. These galaxies benefit from this size threshold, as it helps to reduce the blending effects near the central galaxy. Ideally, the optimal size threshold for each galaxy should be dynamically determined based on its individual radius. However, because the cutout images must be resampled to identical pixel sizes for CNN processing, this approach presents a challenge: the resampled images cannot reflect each galaxy's radius, which is an important morphological feature for photo-$z$ estimation. Therefore, we utilize fixed size threshold in this work, and will investigate the dynamical thresholds and the impact of blendings on photo-$z$ estimations in future research. 

For the convenience of integrating these images into our neural network architecture, we resample the $g$, $r$ and $z$ band images to a resolution of $64\times64$ pixels, while the $W1$ and $W2$ band images are resampled to $32\times32$ pixels, using the Lanczos-3 resampling method, which is consistent with the approach used in the DESI DR9 data release. Note that the images are not resampled to the same pixel sizes, since our neural network model features a forked architecture, comprising two distinct branches that one processes the optical images, and the other handles the near-infrared images. This forked architecture is designed to handle the unique characteristics of the optical and near-infrared images respectively, and can straightforwardly incorporate band sets of other surveys to increase the photo-$z$ performance. 
The optical band images are measured in units of nanomaggies per pixel. However, the WISE band images are initially presented on the Vega magnitude system. We convert the WISE images from Vega to AB magnitude units using the conversion factors recommended by the WISE team, although the specific units of measurements do not affect the neural network performance. 

\subsection{Spectroscopic data} \label{sec: spec data}
To train our deep learning models effectively, we require labels of accurate redshifts. We exclusively use the spectroscopic redshifts provided in the DESI Early Data Release (DESI EDR) for consistency and quality assurance. In an effort to explore potential improvements in accuracy with larger dataset, we intentionally include additional sources from other surveys in Section~\ref{sec: more data}. The selection criteria for spectroscopic sources are strictly defined to ensure the quality and relevance of the data, as follows:
\begin{equation} \label{eq: spec selection}
\begin{split}
    &\qquad\qquad \rm SV\_PRIMARY == True\\
    &\qquad\qquad \rm SPECTYPE == GALAXY\\
    &\qquad\qquad \rm MORPHTYPE\ != PSF\\
    &\qquad\qquad \rm ZWARN == 0\\
    &\qquad\qquad \rm MASKBITS == 0\\
    &\qquad\qquad \rm FLUX\_G, R, Z, W1, W2 > 0\\
    &\qquad\qquad \rm FLUX\_IVAR\_G, R, Z, W1, W2 > 0
\end{split}
\end{equation}
The parameter SV\_PRIMARY indicates the most reliable redshift when multiple measurements are available for the same source. The SPECTYPE is used to confirm the target as a galaxy, aligning with our focus on galaxy data. We select non-PSF sources to maintain consistency with the photometric selection criteria used for imaging data as Equation~\ref{eq: photo selection}. Although some spectroscopically classified galaxies might appear photometrically as PSFs, we exclude these sources from our analysis to ensure data consistency. ZWARN is a flag set by DESI's spectroscopic redshift fitting software, Redrock~\footnote{\url{https://github.com/desihub/redrock}}, where a value of 0 indicates that the spectroscopic redshift is securely determined without issues in the spectrum or the redshift measurement procedure. Other criteria are similar to the ones applied to photometric data mentioned in Section~\ref{sec: image data}.


\subsection{Source categorization}\label{sec: class}
The galaxies from the DESI Legacy Surveys are categorized into four groups, BGS, LRG, ELG and NON, based on DESI's main target selection criteria utilizing optical and near-infrared photometry~\citep{Hahn2023,ZhouRP2023,Raichoor2023}. Among these groups, BGS consist of nearby, luminous galaxies observed during the bright program phase of DESI, designed to probe the local Universe. In contrast, LRG and ELG are primarily observed during DESI's dark time and are essential for tracing the large-scale structure at higher redshifts. LRGs are charaterized as red, massive galaxies that have already ceased their star formation activity. Their spectra display a strong 4000{\AA} break, making their spectroscopic redshifts relatively straightforward to measure~\citep{ZhouRP2020}. This spectral feature also provides a clear signal for estimating photometric redshifts. ELGs targeted by DESI are identified through the prominent [OII] doublet emission lines at $\lambda\lambda$3726,3729 {\AA}. These lines, indicative of high star-formation rates, can be resolved by DESI's spectrograph without the need for a strong continuum, making ELGs excellent tracers for cosmic structure at high redshift~\citep{Raichoor2020}. Additionally, the majority of sources in the DESI Legacy Surveys are not primary targets for spectroscopic observations, likely due to their less distinctive spectroscopic features or unsuitable redshift ranges, falling into the NON group. 

\begin{figure*}
\centering
\includegraphics[width=0.4\linewidth]{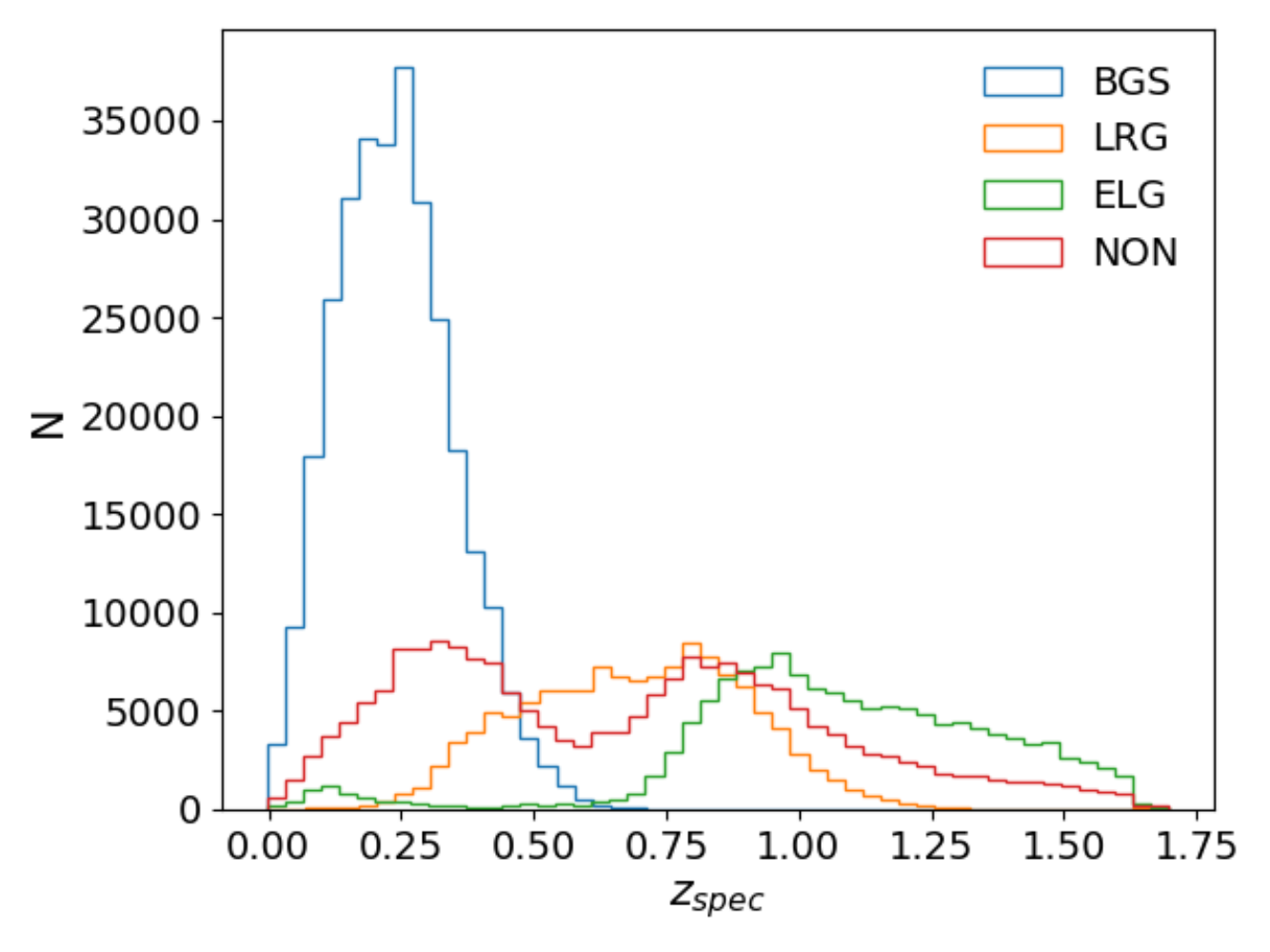}
\includegraphics[width=0.4\linewidth]{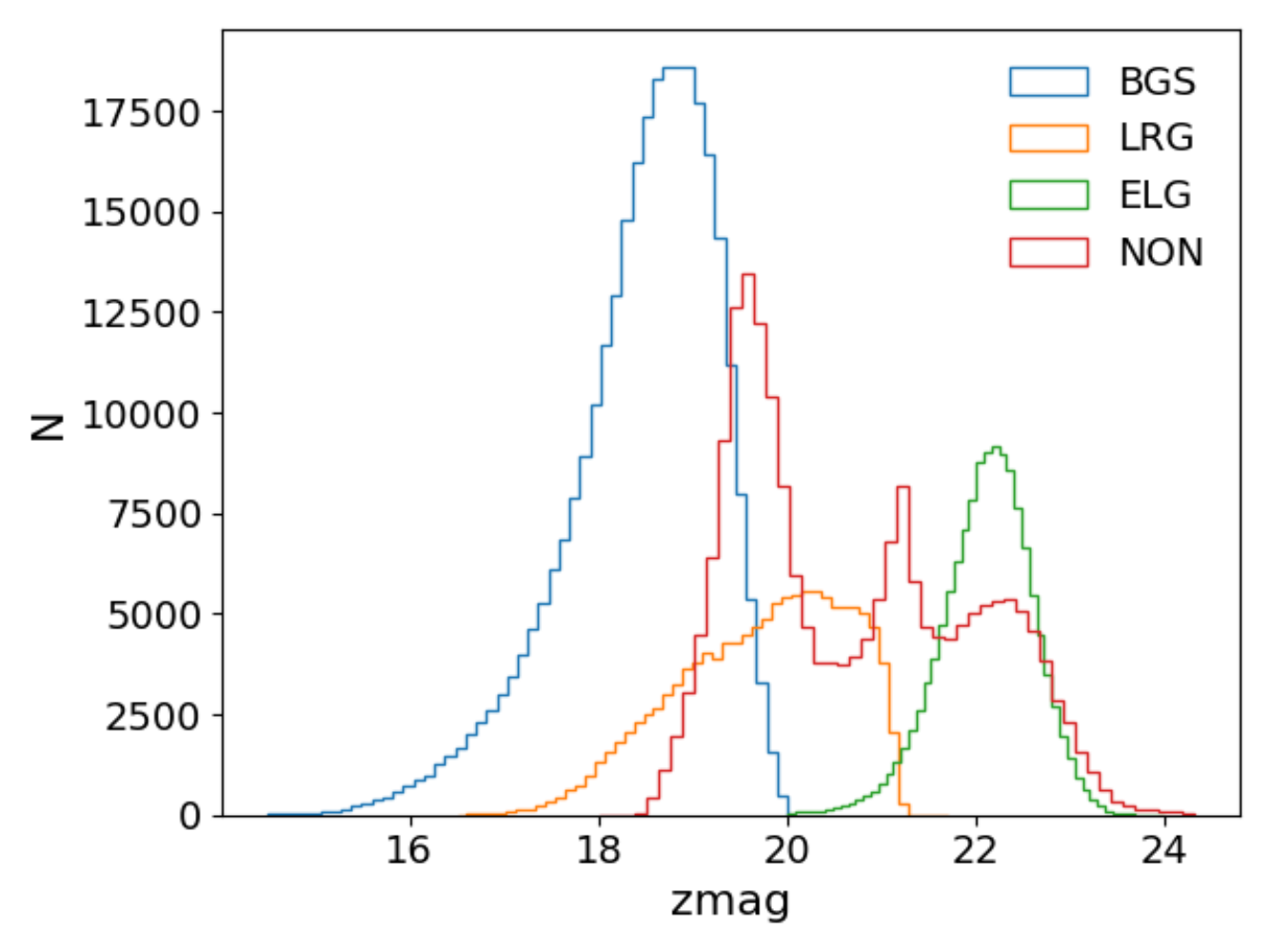}
\caption{\textit{Left}: spectroscopic redshift distribution for BGS, LRG, ELG and NON sources in our study. \textit{Right}: $z$ band magnitude distribution for these sources. We notice that BGS typically includes low-$z$, bright galaixes, while ELG can extend to higher redshift and are fainter. And LRG represent an intermediate group between BGS and ELG. The redshift and $z$ band magnitude distribution for NON both demonstrate double peak, indicating that they are composed of boundary sources of BGS, LRG and ELG. }
\label{fig: z dist spec}
\end{figure*}

The distributions of spectroscopic redshifts and $z$ band magnitudes for BGS, LRG, ELG and NON are displayed in left and right panel of Figure~\ref{fig: z dist spec} respectively. We notice that BGS typically includes low-redshift, luminous galaxies, while ELG can extend up to redshift of $1.6$ and tend to be faint in $z$ band. And LRG represent an intermediate group between BGS and ELG. The redshift and $z$ band magnitude distributions for NON both demonstrate double peak, indicating that this group of sources consists of the boundary sources for BGS, LRG and ELG. 

Given the overlapping redshift ranges for the four groups and the tendency of neural networks to average results across different groups, we decide to estimate the photo-$z$s within each group separately, instead of estimating them collectively that are commonly used in previous studies. This strategy aims to mitigate the potential confusion of sources in feature space, which is critical for enhancing the precision of photometric redshifts. By segregating the groups, we can tailor the model more specifically to the unique characteristics of each group, thereby improving the accuracy of our photo-$z$ estimations. {Note that this concept is not novel. ~\citet{Samui2017} developed a framework called CuBANz~\footnote{\url{https://goo.gl/fpk90V}}, and it divides the data into multiple self-learning clusters based on colors and estimates the photo-$z$ within each cluster. For DESI, the clustering is straightforward using the target selection. However, for other photometric surveys that extend to higher redshifts, determining the optimal clusters requires comprehensive investigations.}

\section{Methodology}\label{sec: methods}
We employ BNN to estimate photometric redshifts along with their uncertainties directly from multi-band images. In this section, we introduce the architecture of BNN, including its CNN backbone and Bayesian layers, training procedure and calibration of uncertainties. The implementation of all networks in our study is carried out using Keras~\footnote{\url{https://keras.io/}} backend on TensorFlow~\footnote{\url{https://www.tensorflow.org/}} and TensorFlow-Probability~\footnote{\url{https://www.tensorflow.org/probability}}.

\subsection{Neural networks}\label{sec: neural network}
\subsubsection{Convolutional neural network}\label{sec: point estimate backbone}
Estimating photometric redshifts directly from multi-band photometric imaging data offers several advantages, particularly by circumventing potential inaccuracies induced by photometric measurements. Additionally, this method allows for the natural integration of morphological information, which can significantly enhance photo-$z$ performance. Since a decreasing trend exists in effective radius with increasing redshift, this morphological feature can help resolve degeneracy between sources at low and high redshifts, thereby reducing the outlier fraction for photo-$z$ estimation~\citep{2018MNRAS.475.3613S, 2021AIPC.2319d0002S}. To leverage these advantages, we employ convolutional neural networks (CNNs), which are exceptionally suited for image processing tasks. 

Our CNN for estimating photometric redshifts employs a fork-like architecture designed to process galaxy images in two different pixel sizes, as detailed in Section~\ref{sec: image data}. This architecture comprises two separate branches: one for the $grz$ band images and another for the $W1W2$ ones. These branches are designed to handle the unique characteristics of the optical and near-infrared data respectively, before their learned features are concatenated for further analysis.

The initial layer in each branch is a convolutional layer with a kernel size of 7, designed to extract a primary feature map from the input images. Following this, we apply the Convolutional Block Attention Module (CBAM; ~\citet{2018arXiv180706521W}), which enhances the network's focus on informative features by integrating attention mechanisms in both spatial and channel dimensions. This module is lightweight and can be seamlessly integrated into any CNN architecture, improving performance by focusing the network's attention on salient features. Each branch then continues with a series of ResNet blocks, 16 blocks for the $grz$ branch and 12 for the $W1W2$ branch. These ResNet blocks can help mitigate the problem of vanishing gradients, a common issue in deep neural networks~\citep{2015arXiv151203385H}. The feature maps are progressively refined to spatial dimension of $2\times2$. After passing through the series of ResNet blocks, global average pooling is applied to each branch to vectorize the feature maps into vectors. 

The vectors from each branch are then concatenated, and the combined vector is processed through two fully connected layers to derive the final output. To ensure effective learning and generalization, we incorporate the ReLU activation function~\citep{2018arXiv180308375A} and batch normalization~\citep{2015arXiv150203167I} after each convolutional and fully connected layer. ReLU aims to implement non-linearity, while the batch normalization helps to reduce overfitting and improve the overall performance of the network.

\subsubsection{Bayesian neural networks}\label{sec: BNN}
In astronomical and cosmological studies, quantifying the uncertainty of predictions is crucial. {Certain works transform the regression problem into a classification problem to obtain the probability distribution functions (PDFs)~\citep{Pasquet2019,Treyer2024}. Others employ the framework of Bayesian neural networks (BNNs). In this study, we adopt the latter approach. The uncertainties associated with predictions from neural network can be categorized into two distinct components: epistemic uncertainty and aleatoric uncertainty. Epistemic uncertainty originates from the model itself, and it is commonly addressed by employing multiple networks with varying configurations. The results from these networks are then averaged to determine the final output, thereby incorporating the uncertainty associated with the model. This approach introduces the concept of a Bayesian neural network (BNN), which utilizes probabilistic distributions to represent weights. Consequently, each weight sample provides a different network configuration. While this method effectively addresses epistemic uncertainty, it is equally crucial to acknowledge the presence of aleatoric uncertainty, which arises from the inherent variability of the data.} One effective way to incorporate aleatoric uncertainty is through a Mixture Density Network (MDN), which outputs a combination of several distributions, as detailed by~\citet{Bishop1994}. For many applications, a single Gaussian distribution is sufficient. A probabilistic Bayesian neural network combines these approaches, capturing both epistemic and aleatoric uncertainties. This type of BNN effectively addresses the complexities of uncertainty in model predictions. Please refer to~\citet{2020PhRvD.102j3509H} and ~\cite{Zhou2022pho} for more details for this category of network. 

We construct our Bayesian model using transfer learning, a technique where a model trained on one task is adapted to improve performance on a related one. In our setup, we retain the architecture of the CNN up to the fully connected layers as the backbone for the BNN, with the weights remaining fixed. Here we attempt two categories of Bayesian layers, specifically Monte Carlo dropout (MC-dropout)~\citep{2015arXiv150602142G} or Multiplicative Normalizing Flows (MNFs)~\citep{2017arXiv170301961L}, appending at the end of the backbone network. Unlike standard dropout which is only active during training~\citep{JMLR:v15:srivastava14a}, MC-dropout functions during testing as well, naturally offering a method to simulate various network configurations through random weight dropout. MNF, on the other hand, transforms simple Gaussian weight distributions into more complex forms using normalizing flows~\citep{2015arXiv150505770J}, thus enhancing the robustness of BNN. 

In our implementation, we integrate two layers of MC-dropout or MNF, with network's output modeled as a Gaussian distribution to capture aleatoric uncertainty -- a reasonable assumption for photo-$z$ of each source. {Note that other distributions can also be utilized. However, for the simplicity, we will exclusively consider the Gaussian distribution. The analysis and applications of other distributions will be explored in future work.} Contrary to approach mentioned in~\citet{Zhou2022pho}, where all weight layers are replaced with Bayesian ones, we only append limited Bayesian layers to the trained backbone. This strategy can significantly restrict the model complexity and speed up the training process. 

\subsection{Training}\label{sec: training}
We train four separate networks for BGS, LRG, ELG and NON to avoid potential performance degradation that could occur from training these sources collectively due to their overlapping redshift ranges and the averaging effect of networks, as detailed in Section~\ref{sec: class}. For comparison, we also train these sources collectively using one network.

Given the nature of the target selection, the redshift distributions for BGS and LRG exhibit long tails that extend far beyond the typical redshift range of interest for DESI. The photo-$z$ estimation can be problematic at redshifts where the sources are rare. And same situation occurs at low redshift end for LRG and ELG sources. To mitigate these issues, we limit the dataset of each target by including sources at spec-$z$s within 0.3\% to 99.7\% quantiles. Note that this limitation cannot fully address the low redshift contaminants produced in the ELG selection. These contaminants are technically not ELG population DESI aims to observe in main survey as the resolving of their [OII] doublets are difficult because of the decreasing resolution towards the blue end of instrument's wavelength coverage~\citep{DESIcoll2016}. Nonetheless, these are included as ELG in our analysis since they meet the target selection criteria and have secure spectroscopic redshifts.

And then we split the dataset of each target into training, validation and testing as a ratio of 7:1:2. To reduce overfitting, we augment the training data by including rotated and mirrored versions of the images, and another version by introducing a small amount of Gaussian noise (mean = $0$, variance = $\rm 1E-6$) to each image. This level of noise does not impact the signal-to-noise ratio, photometric measurements, or morphology of the galaxies, but it does alter pixel values slightly, which can be detected by the deep learning model. This increases both the size of the dataset and the model's robustness against adversarial attacks~\citep{qiu2019}. In summary, the training size is augmented by 9 $\times$ original size.

The training process for the CNN is straightforward. We use Huber loss~\citep{10.1214/aoms/1177703732}, which is less sensitive to outliers than mean squared error (MSE), making it more suitable for robust regression. The network is optimized using Adam optimizer~\citep{kingma2017adam} and the model with lowest validation loss value is preserved using the ModelCheckpoint callback to serve as the backbone for the Bayesian model. For the BNNs, we optimize using the negative log-likelihood function. The backbone weights transferred from the CNN are set as untrainable to preserve the learned features. The dropout rate for the MC-dropout is a critical hyperparameter, experimented at rates of 0.01, 0.1 and 0.5, with the optimal rate chosen for the final model. As for the MNF model, the parameters are left at their default settings, which adapts 50 layers for masked RealNVP normalizing flow~\citep{2015arXiv150203509G, 2016arXiv160508803D}. Similarly, the model with lowest validation loss is preserved as our final BNN model for further investigation on testing data.

\subsection{Calibration}\label{sec: calibration}
The uncertainties derived from Bayesian neural networks (BNN) must adhere to statistical principles, ensuring that the true values of $x\%$ of the samples lie within the corresponding $x\%$ confidence intervals~\citep{2017ApJ...850L...7P, 2020PhRvD.102j3509H}. This concept is crucial for confirming that the uncertainties are properly calibrated. To assess this, we can use the reliability diagram, which plots the coverage probability against the confidence interval. Ideally, this diagram should exhibit a straight diagonal line, indicating that the uncertainties are well calibrated. Deviations from this line suggest a need for recalibration.

Although recalibration of BNNs can be done by adjusting their hyperparameters, such an approach typically requires repetitive retraining, making it a time-intensive process. Therefore, post-training calibration is often a more feasible option. Techniques for this method have been discussed extensively in~\citet{2020PhRvD.102j3509H}. In our study, we implement a straightforward Beta calibration method, as introduced in~\citet{Kull2017}. This method involves scaling all uncertainties estimated by the BNN by a factor to adjust the reliability diagram towards diagonal line. This scaling ensures the calibrated uncertainties more accurately reflect the true confidence intervals, enhancing the reliability and trustworthiness of our photometric redshift estimations. 

\section{Results}\label{sec: results}
In this section, we demonstrate the accuracy that can be achieved for each category of sources by CNN and BNN, and compare them with the results from previous studies. Finally, we introduce our photo-$z$ catalogue for DESI legacy imaging surveys. 

\subsection{Results of CNN} \label{sec: point estimate results}

\begin{table*}
\caption{{The performance of CNN for BGS, LRG, ELG and NON sources in our work using separate and collective training strategy. The results in ~\citet{Zou2019} are shown for comparison. $\eta$, $\sigma_{\rm NMAD}$ and $\overline{\Delta z}$ indicate the outlier percentage, accuracy and mean bias respectively, and $N_{\rm training}$ is the approximate size of the training sets in units of million. Additionally, the point estimates from BNN under separate training strategy are also illustrated.}}
\label{tab: compare}
\hspace*{-1.5cm}
\begin{tabular}{|c|cccc||cccc|}
\hline
Targets                          & \multicolumn{4}{c||}{BGS}                                                                                                                  & \multicolumn{4}{c|}{LRG}                                                                                                                 \\ \hline
Metrics                          & \multicolumn{1}{c|}{$\eta$}  & \multicolumn{1}{c|}{$\sigma_{\rm NMAD}$} & \multicolumn{1}{c|}{$\overline{\Delta z}$} & $N_{\rm training}$ & \multicolumn{1}{c|}{$\eta$} & \multicolumn{1}{c|}{$\sigma_{\rm NMAD}$} & \multicolumn{1}{c|}{$\overline{\Delta z}$} & $N_{\rm training}$ \\ \hline
Our work (Separately)            & \multicolumn{1}{c|}{0.14\%}  & \multicolumn{1}{c|}{0.020}               & \multicolumn{1}{c|}{-0.0031}               & 0.3 m              & \multicolumn{1}{c|}{0.68\%} & \multicolumn{1}{c|}{0.030}               & \multicolumn{1}{c|}{0.0111}                & 0.1 m              \\ \hline
Our work (Bayesian)              & \multicolumn{1}{c|}{0.14\%}  & \multicolumn{1}{c|}{0.018}               & \multicolumn{1}{c|}{-0.0015}               & 0.3 m              & \multicolumn{1}{c|}{0.45\%} & \multicolumn{1}{c|}{0.026}               & \multicolumn{1}{c|}{0.0022}                & 0.1 m              \\ \hline
Our work (Collectively)          & \multicolumn{1}{c|}{0.83\%}  & \multicolumn{1}{c|}{0.029}               & \multicolumn{1}{c|}{0.0160}                & -                  & \multicolumn{1}{c|}{1.07\%} & \multicolumn{1}{c|}{0.033}               & \multicolumn{1}{c|}{0.0151}                & -                  \\ \hline
\citet{Zou2019} & \multicolumn{1}{c|}{0.19\%}  & \multicolumn{1}{c|}{0.013}               & \multicolumn{1}{c|}{0.0}                   & 1.2 m              & \multicolumn{1}{c|}{0.18\%} & \multicolumn{1}{c|}{0.016}               & \multicolumn{1}{c|}{0.0013}                & 0.8 m              \\ \hline \hline
Targets                          & \multicolumn{4}{c||}{ELG}                                                                                                                  & \multicolumn{4}{c|}{NON}                                                                                                                 \\ \hline
Metrics                          & \multicolumn{1}{c|}{$\eta$}  & \multicolumn{1}{c|}{$\sigma_{\rm NMAD}$} & \multicolumn{1}{c|}{$\overline{\Delta z}$} & $N_{\rm training}$ & \multicolumn{1}{c|}{$\eta$} & \multicolumn{1}{c|}{$\sigma_{\rm NMAD}$} & \multicolumn{1}{c|}{$\overline{\Delta z}$} & $N_{\rm training}$ \\ \hline
Our work (Separately)            & \multicolumn{1}{c|}{16.65\%} & \multicolumn{1}{c|}{0.112}               & \multicolumn{1}{c|}{0.0257}                & 0.1 m              & \multicolumn{1}{c|}{9.44\%} & \multicolumn{1}{c|}{0.058}               & \multicolumn{1}{c|}{0.0014}                & 0.2 m              \\ \hline
Our work (Bayesian)              & \multicolumn{1}{c|}{16.07\%} & \multicolumn{1}{c|}{0.107}               & \multicolumn{1}{c|}{0.0184}                & 0.1 m              & \multicolumn{1}{c|}{7.87\%} & \multicolumn{1}{c|}{0.052}               & \multicolumn{1}{c|}{0.0140}                & 0.2 m              \\ \hline
Our work (Collectively)          & \multicolumn{1}{c|}{15.78\%} & \multicolumn{1}{c|}{0.108}               & \multicolumn{1}{c|}{0.0154}                & -                  & \multicolumn{1}{c|}{8.15\%} & \multicolumn{1}{c|}{0.053}               & \multicolumn{1}{c|}{0.0304}                & -                  \\ \hline
\citet{Zou2019} & \multicolumn{1}{c|}{6.23\%}  & \multicolumn{1}{c|}{0.053}               & \multicolumn{1}{c|}{0.0067}                & 0.02 m             & \multicolumn{1}{c|}{2.46\%} & \multicolumn{1}{c|}{0.024}               & \multicolumn{1}{c|}{0.0022}                & 0.6 m              \\ \hline
\end{tabular}
\end{table*}

\begin{figure*}
    \centering
    \includegraphics[width=0.4\textwidth]{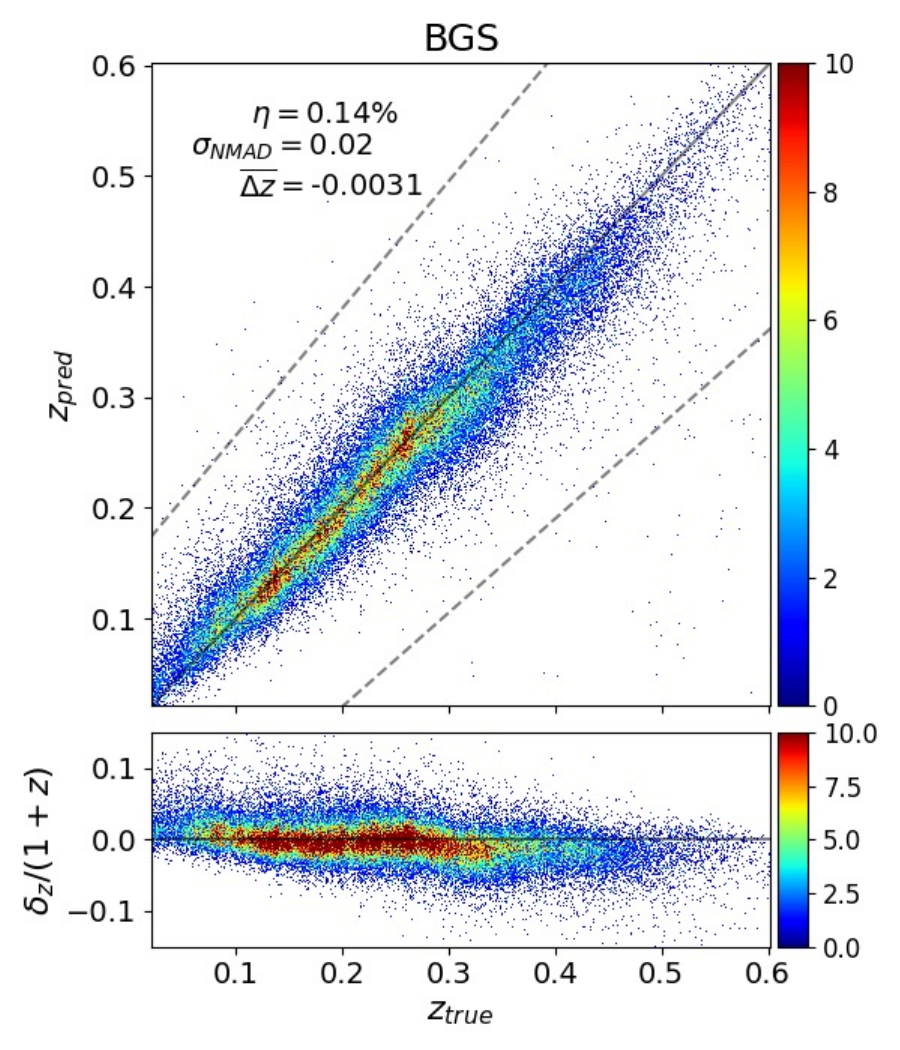}
    \includegraphics[width=0.4\textwidth]{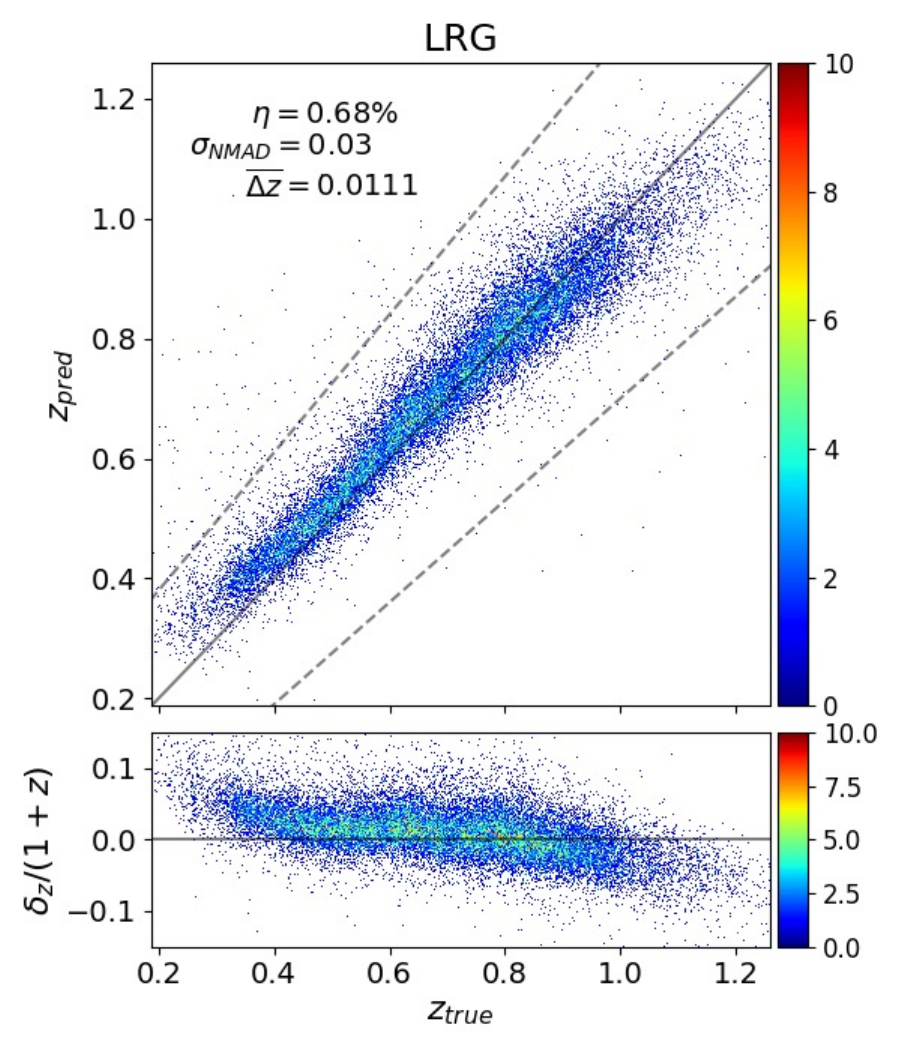}\\
    \includegraphics[width=0.4\textwidth]{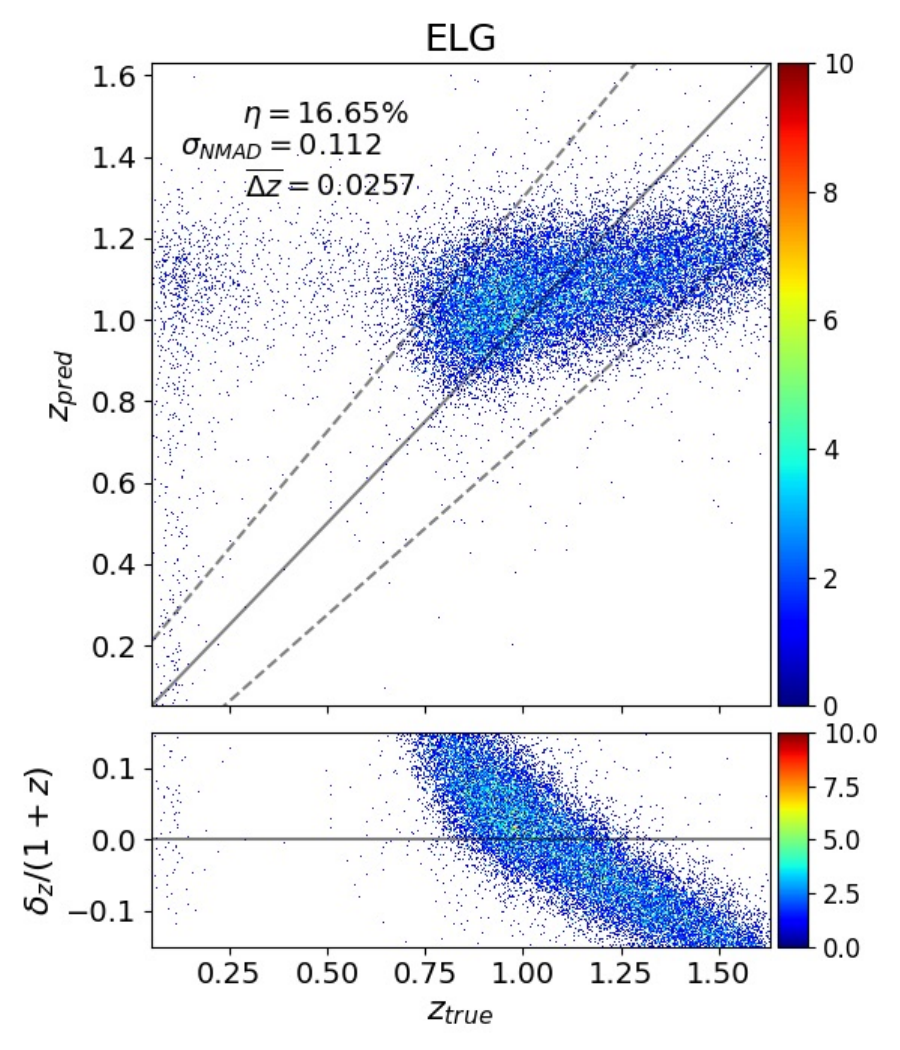}
    \includegraphics[width=0.4\textwidth]{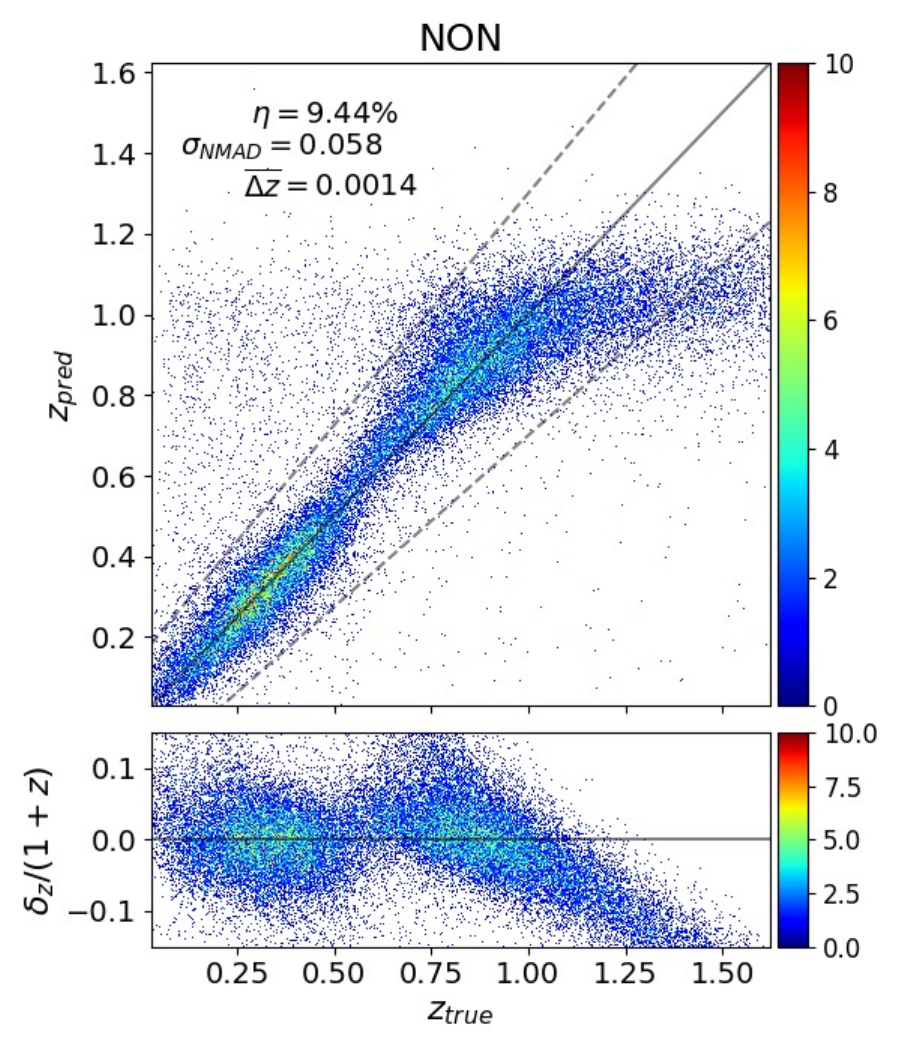}
    \caption{Results of CNN for BGS, LRG, ELG and NON targets for testing sets. $\eta$, $\sigma_{\rm NMAD}$ and $\overline{\Delta z}$ indicates outlier percentage, accuracy and mean bias respectively. {The black solid line represents the one-to-one correspondence, while the black dashed line indicates an outlier of $|\Delta z| > 0.15(1 + z_{\rm true})$. Additionally, the color bar suggesting the density of sources per pixel in plot is also illustrated.}}
    \label{fig: point estimates}
\end{figure*}

\begin{figure}
    \centering
    \includegraphics[width=0.5\textwidth]{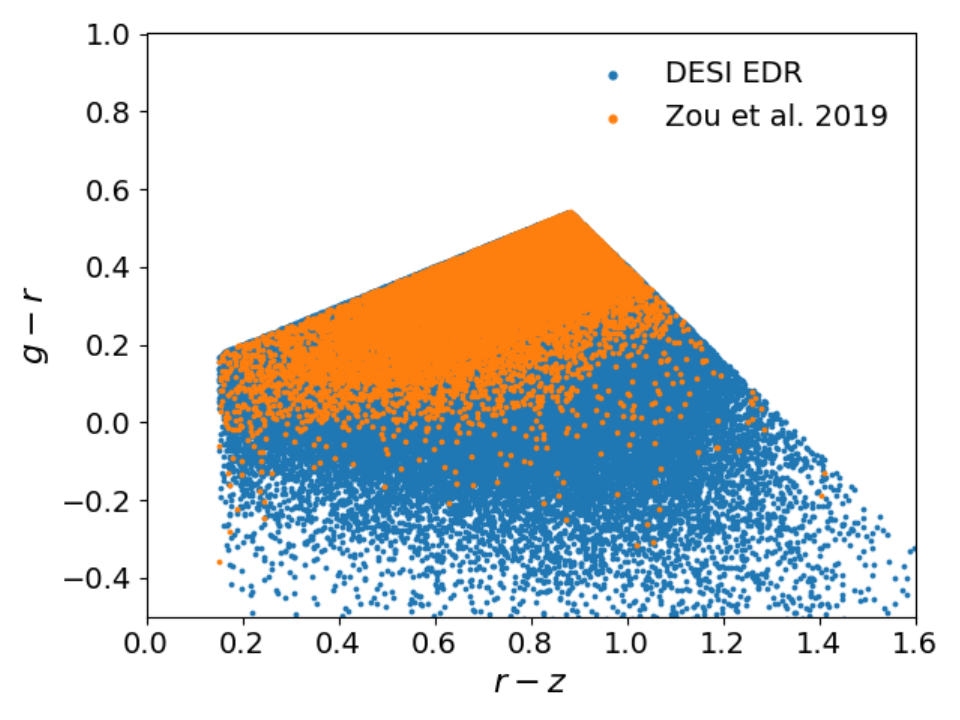}
    \caption{Color coverage comparison for ELG between our work and~\citet{Zou2019}. The distinction for the sources explains the much worse performance for ELG than their work.}
    \label{fig: elg color}
\end{figure}

The accuracy of CNN critically influences the performance of uncertainty estimations in BNN, as the former serves as the backbone of Bayesian models. {We employ three metrics to evaluate our models, outlier percentage $\eta$, accuracy $\sigma_{\rm NMAD}$ and  mean bias $\overline{\Delta z}$, defined as follows:}
\begin{equation}\label{eq: eta}
    \eta = \frac{N_{|\Delta z|/(1 + z_{\rm true}) > 0.15}}{N_{\rm total}},
\end{equation}
\begin{equation}\label{eq: sigma}
    \sigma_{\rm NMAD} = 1.48\times{\rm median}\left(\left|\frac{\Delta z - {\rm median}(\Delta z)}{1 + z_{\rm true}} \right|\right),
\end{equation}
{
\begin{equation}\label{eq: bias}
    \overline{\Delta z} = \frac{\sum \Delta z / (1 + z_{\rm true})}{N_{\rm total}},
\end{equation}
}
where $\Delta z = z_{\rm pred} - z_{\rm true}$, with $z_{\rm pred}$ and $z_{\rm true}$ representing the predicted photometric and true spectroscopic redshifts respectively. {The outlier percentage, $\eta$, $\sigma_{\rm NMAD}$ and $\overline{\Delta z}$ quantifies the fraction of sources with severely inaccurate redshift estimations, the accuracy that is robust against outliers and the mean residual of predictions, respectively.}

The CNN results for BGS, LRG, ELG and NON sources using separate training strategy are illustrated in Figure~\ref{fig: point estimates}. {Directly deriving photo-$z$ from galaxy images yields $\eta=0.14\%$, $\sigma_{\rm NMAD}=0.020$ and $\overline{\Delta z}=-0.0031$ for BGS, and $\eta=0.68\%$, $\sigma_{\rm NMAD}=0.030$ and $\overline{\Delta z}=0.0111$ for LRG, with training datasets of approximately 0.3 million and 0.1 million sources, respectively.} {However, for ELG, the weak and biased correlation between spec-$z$ and photo-$z$ results in $\eta=16.65\%$, $\sigma_{\rm NMAD}=0.112$ and $\overline{\Delta z}=0.0257$, indicating significant challenges in accurately estimating redshifts for this group.} {As for NON, we notice that this group can achieve $\eta=9.44\%$, $\sigma_{\rm NMAD}=0.058$ and $\overline{\Delta z}=0.0014$, between the performance of BGS, LRG and ELG.} The results of NON at low redshift are relatively reasonable, while at high redshift, the correlation and bias become similar to the ELG group. To more explicitly explain the distinct behaviors, we use unsupervised learning technique to further analyze these four groups in Section~\ref{sec: umap}. 

The results using separate and collective training strategy are outlined in Table~\ref{tab: compare}. The performance for BGS and LRG significantly improves under separate strategy compared to the collective approach, while ELG and NON sources show contrary results, exhibiting slightly better performance when estimated collectively. This improvement for ELG and NON sources can be attributed to the expanding data sizes for these two categories of sources when combined together. This outcome underscores the advantages of categorizing sources based on their characteristics for photo-$z$ estimations. By categorizing the source types for training, the models can more accurately learn the specific features and redshift distributions unique to each group, enhancing the precision of photo-$z$ estimations. Furthermore, this approach helps to avoid the averaging effect that can dilute the accuracy when distinct source types are estimated collectively. Therefore, the source categorization and separate training strategy are beneficial for optimizing the photo-$z$ estimations.

Additionally, in Table~\ref{tab: compare}, we compare our results to those in ~\citet{Zou2019}, who used spec-$z$s from multiple surveys like SDSS~\citep{2018ApJS..235...42A}, VVDS~\citep{2005A&A...439..845L,2008A&A...486..683G} and zCOSMOS~\citep{2007ApJS..172...70L} to derive photo-$z$s for DESI sources through a local linear regression algorithm based on photometric measurements. Unlike our approach, which trains models separately for each group of sources, their method trains all sources collectively. {Under the same metrics defined previously in Equation~\ref{eq: eta}, Equation~\ref{eq: sigma} and Equation~\ref{eq: bias}, our BGS model shows slightly better performance in outlier percentage, albeit with a worse $\sigma_{\rm NMAD}$, using fewer training sources.} For LRG, our results are not comparable due to our much smaller dataset, but expanding our training set with additional LRG data from their study indeed improves our performance, reaching similar or even better outcomes, as discussed in Section~\ref{sec: more data}. The training data in their work predominantly come from SDSS observations, which primarily target BGS and LRG. Consequently, the color coverage of $r-z$ vs. $g-r$ is not as comprehensive as that of DESI, as shown in Figure~\ref{fig: elg color}, leading to non-comparable performance for ELG between our study and theirs. Moreover, the insufficient training data and incomplete color coverage jointly attribute to the poor accuracy of NON sources in our study.

\begin{figure*}
    \centering
    \includegraphics[width=0.4\textwidth]{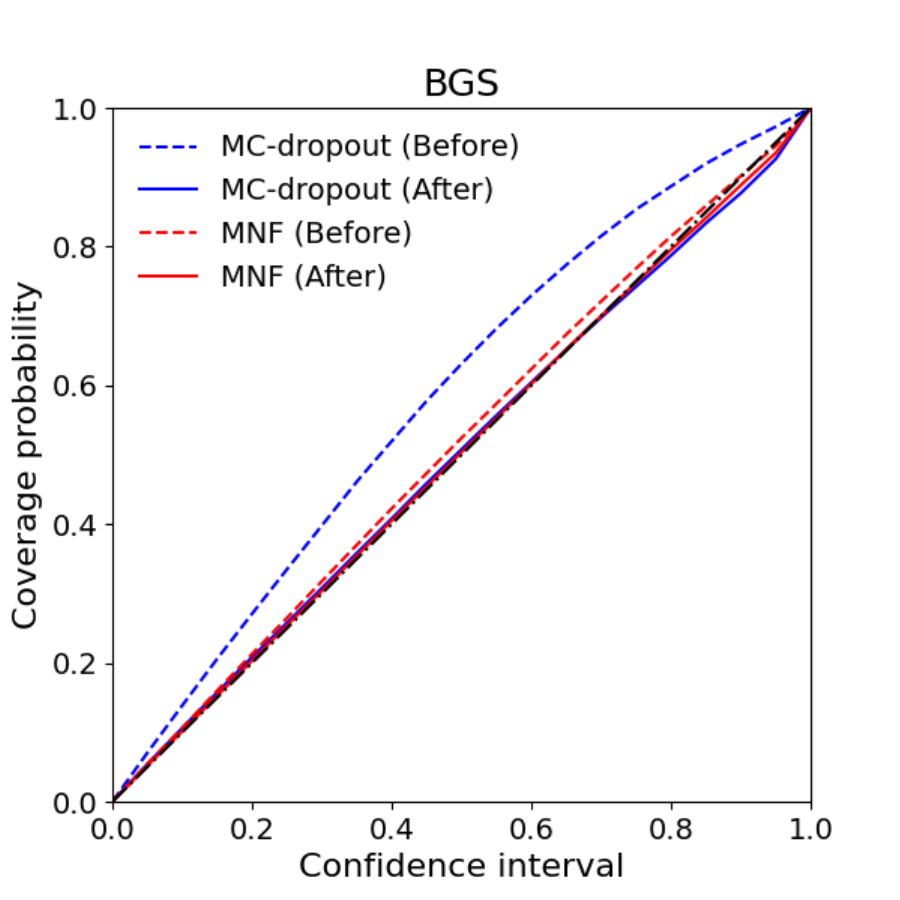}
    \includegraphics[width=0.4\textwidth]{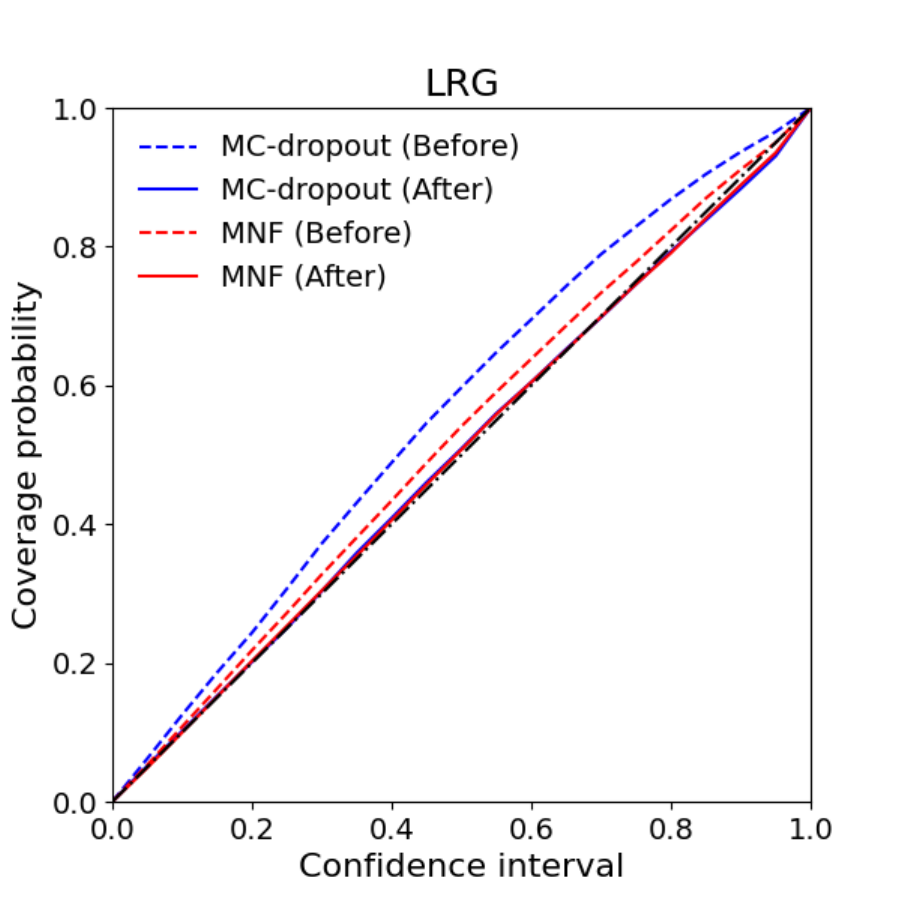}\\
    \includegraphics[width=0.4\textwidth]{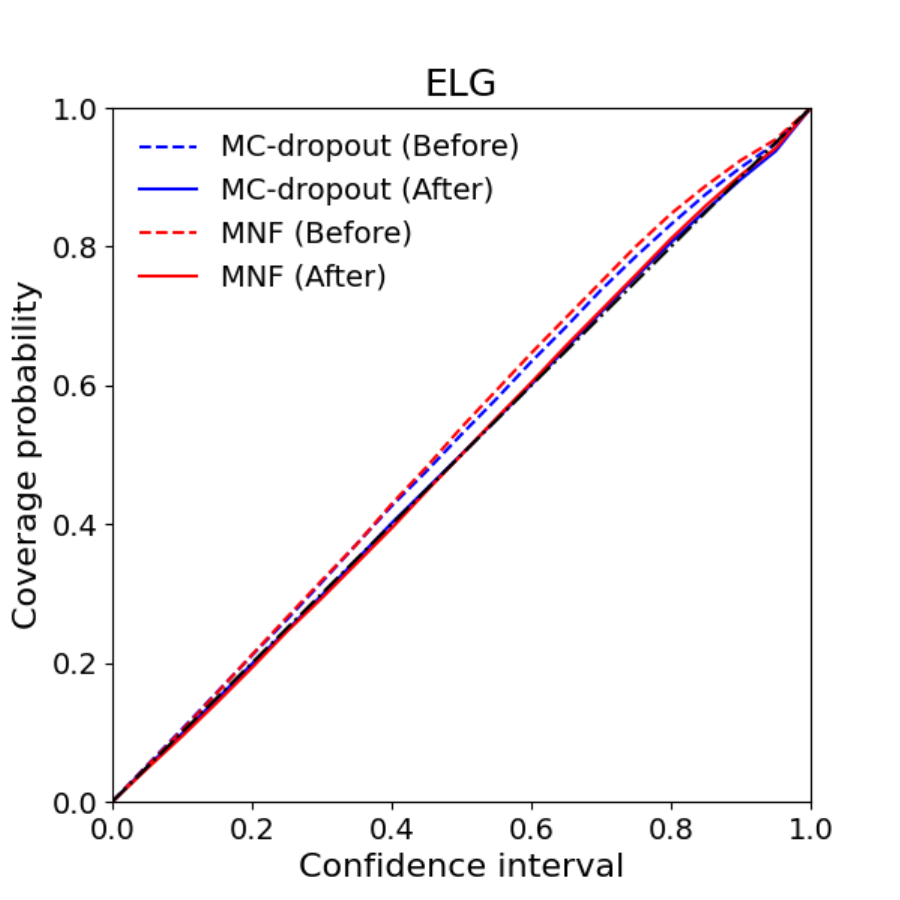}
    \includegraphics[width=0.4\textwidth]{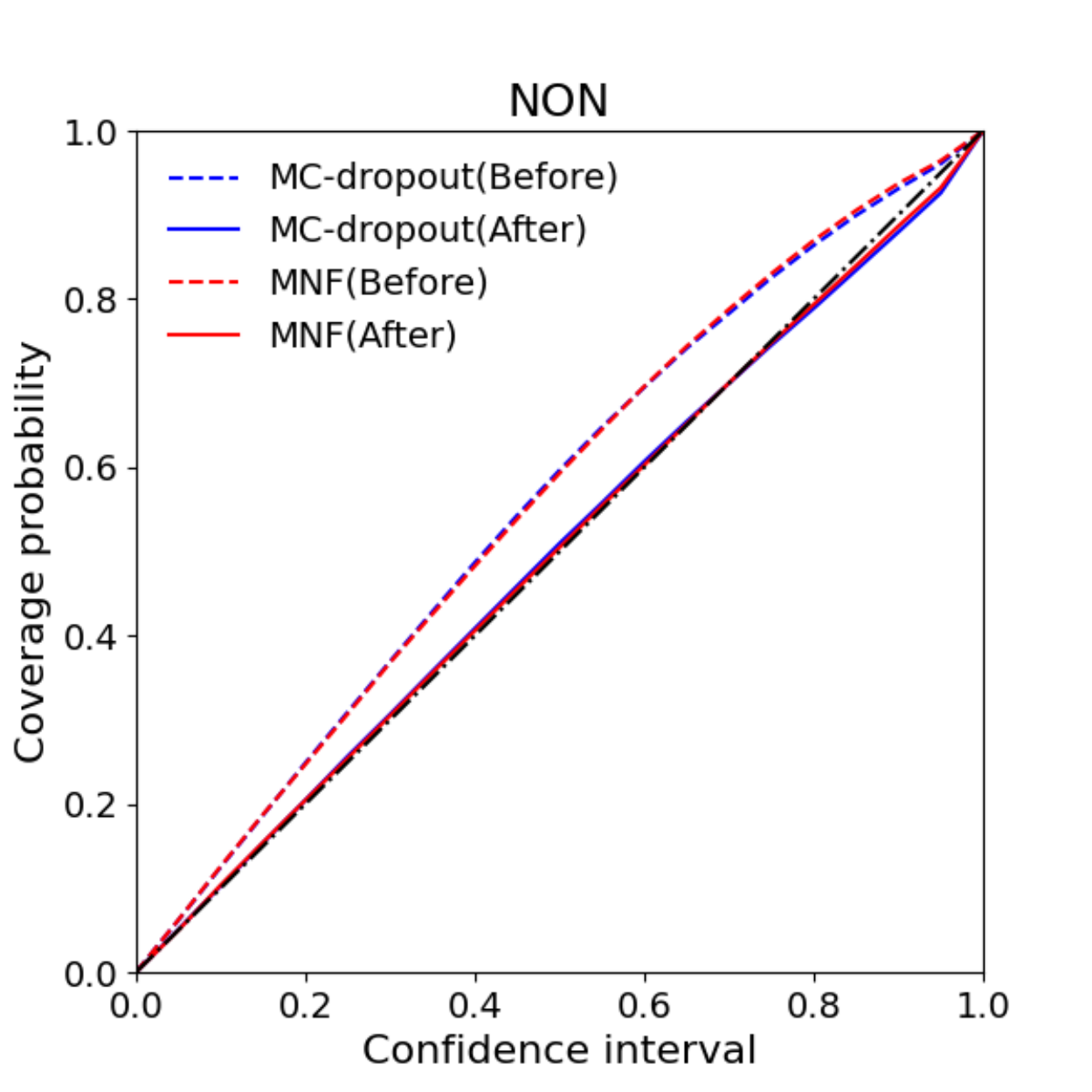}
    \caption{Reliability diagram for BGS, LRG, ELG and NON sources using MC-dropout and MNF Bayesian model respectively. We notice that the MNF models for BGS and LRG are almost self-calibrated, on the contrary, MC-dropout overestimates the uncertainties. Employing the Beta calibration method, the statistical principle is effectively followed.}
    \label{fig: reliability diagram}
\end{figure*}

\subsection{Results of BNN}\label{sec: BNN results}

\begin{figure*}
    \centering
    \includegraphics[width=0.4\textwidth]{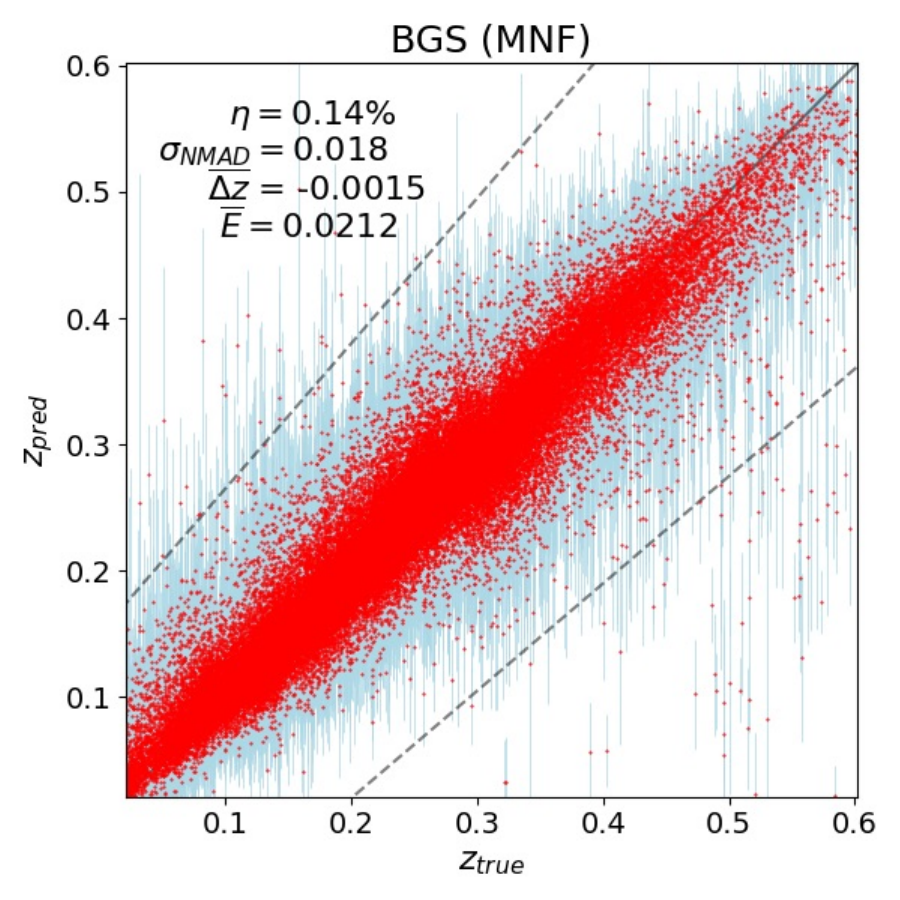}
    \includegraphics[width=0.4\textwidth]{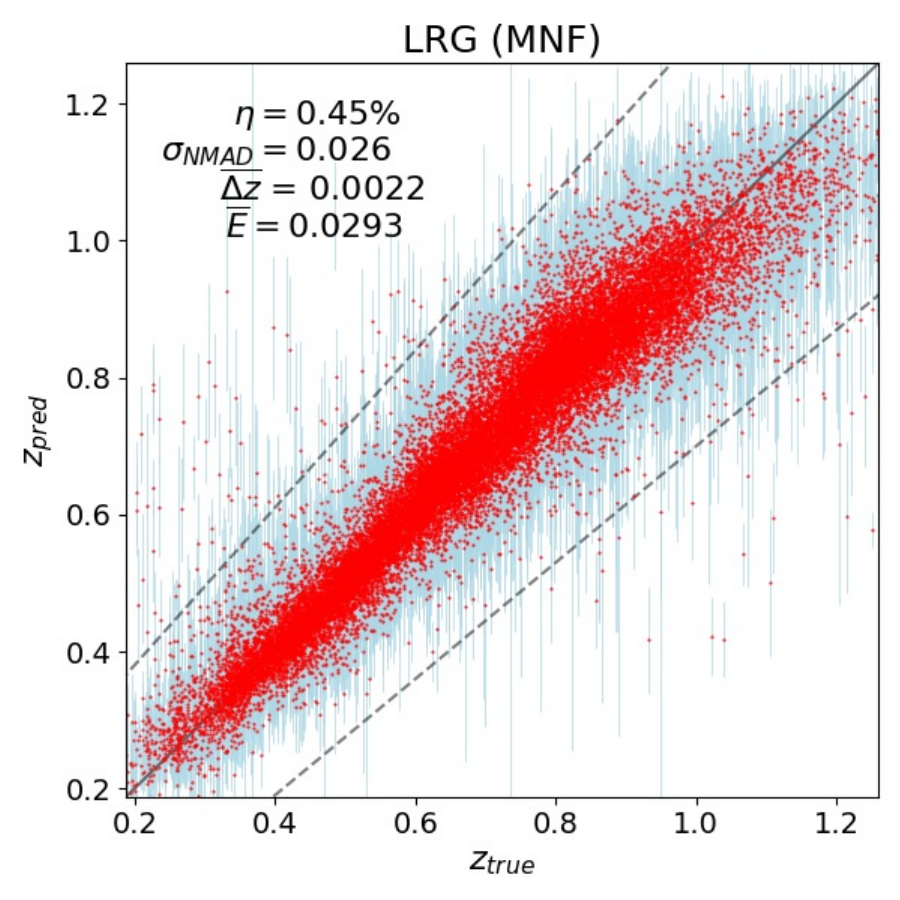}\\
    \includegraphics[width=0.4\textwidth]{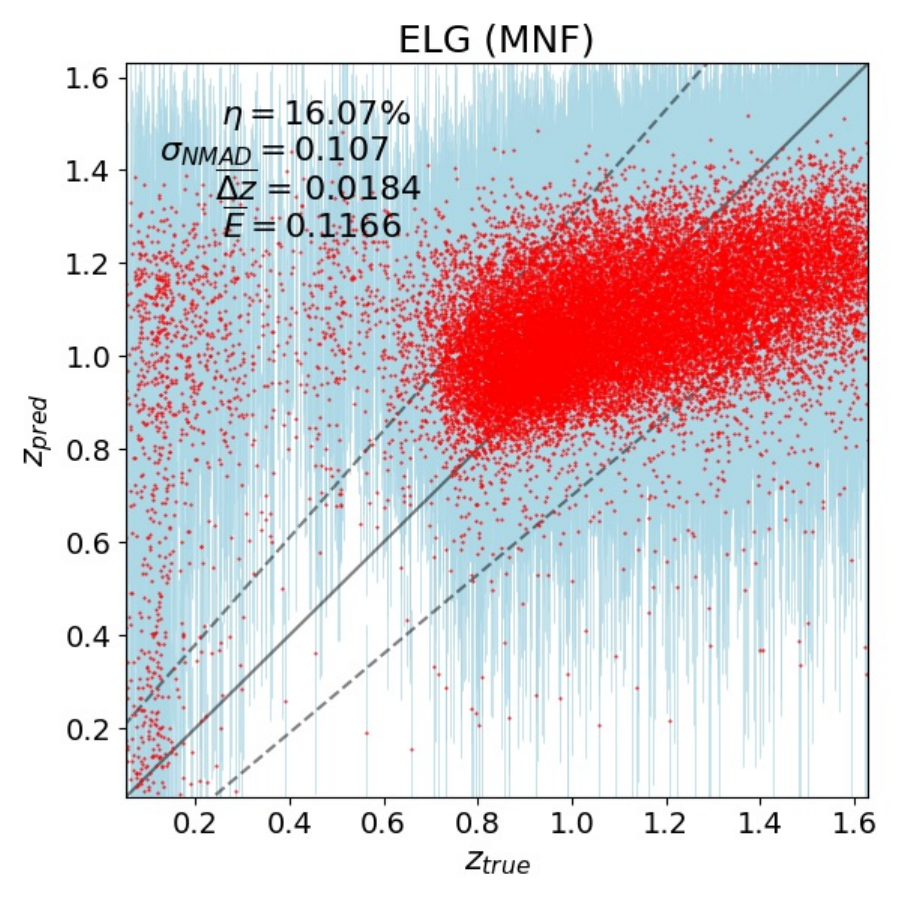}
    \includegraphics[width=0.4\textwidth]{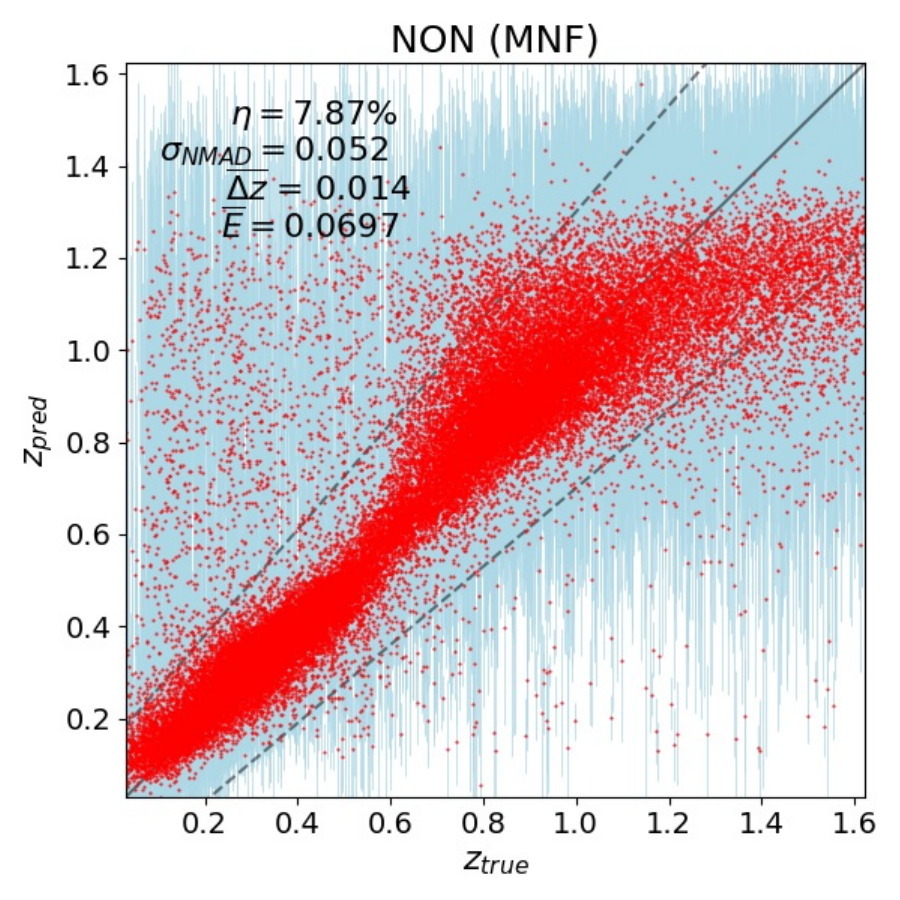}
    \caption{{Results for BGS, LRG, ELG and NON sources using MNF Bayesian models. The uncertainties for each source is indicated by lightblue bar. $\eta$, $\sigma_{\rm NMAD}$, $\overline{\Delta z}$ and $\overline{E}$ suggest outlier percentage, accuracy, mean bias and mean uncertainty respectively. The black solid line represents the one-to-one correspondence, while the black dashed line indicates the outlier of $|\Delta z|>0.15(1 + z_{\rm true})$.
    }}
    \label{fig: mnf results}
\end{figure*}

In Bayesian neural networks, varying network configurations and output distributions are used to capture epistemic and aleatoric uncertainties respectively. To achieve this, the trained network is repeatedly sampled with the testing data; in our study, we conduct this sampling procedure for 200 times. We employ two Bayesian architectures: MC-dropout and MNF. For MC-dropout, we experiment on dropout rate of $0.01$, $0.1$ and $0.5$, finding that the choice of rate have negligible impact on results after a sufficient optimization period, in our case, $100$ epochs. Thus, we select a dropout rate of 0.01 for our MC-dropout model. The MNF model, using default settings, yields results comparable to those of the MC-dropout approach. 

Both models' uncertainty estimates are calibrated employing Beta calibration technique as described in Section~\ref{sec: calibration}. The reliability diagrams for BGS, LRG, ELG and NON before and after calibration are displayed in Figure~\ref{fig: reliability diagram}. We notice that the MNF models tend to be almost self-calibrated compared to MC-dropout models. This outcome suggests that trainable weights, represented by complex distributions derived from Gaussian distributions through normalizing flows, are more effective than merely altering network configurations with dropout layers. 

{Here we introduce mean uncertainty to assess the uncertainty estimations, defined as:
\begin{equation}
    \label{eq:weighted_mean_uncertainty}
    \overline{E} = \frac{\sum E/(1 + z_{\rm{true}})}{N_{\rm total}}, 
\end{equation}
where $E$ is the uncertainty. Commonly, more robust Bayesian model will produce lower mean uncertainty measurement.} The performance of the MNF and MC-dropout models, in terms of both point and uncertainty estimations, is almost identical, as shown in Figure~\ref{fig: mnf results} and Figure~\ref{fig: dropout results} in Appendix~\ref{app: dropout}. Moreover, the point estimation metrics for these models are comparable or even superior to those achieved by CNN models shown in Figure~\ref{fig: point estimates}, indicating that the addition of Bayesian layers can further optimize the performance. Given the results observed in the reliability diagrams for calibration, we ultimately select the MNF model as our preferred model for creating the photo-$z$ catalogue for DESI sources. 

\subsection{Photo-$z$ catalogue}\label{sec:catalog}

\begin{table}
\caption{The counts of sources in our photometric redshift catalogue across different groups in northern and southern hemisphere. Note that the numbers in Summary column do not exactly equal summation of all groups, since there are some overlaps between BGS and LRG groups.}
\label{tab: sources}
\begin{tabular}{|c|c|c|c|c|}
\hline
Targets & BGS        & LRG        & NON         & Summary     \\ \hline\hline
North   & 5,943,940  & 3,155,060  & 35,217,691  & 43,622,109  \\ \hline
South   & 18,068,603 & 9,291,046  & 111,015,601 & 136,154,439 \\ \hline
Total   & 24,012,543 & 12,446,106 & 146,233,292 & 179,776,548 \\ \hline
\end{tabular}
\end{table}

Our photometric redshift catalogue includes $\sim0.18$ billion sources totally. Among these sources, the number of BGS, LRG and NON are 24 million, 12 million and 0.15 billion respectively. ELGs are excluded considering their poor performance. To improve the precision, NON sources are constraint by $z < 21.3$, as discussed in Section~\ref{sec: umap}. The exact numbers of these sources in northern and southern hemisphere are demonstrated in Table~\ref{tab: sources}. Note that there are overlaps between BGS and LRG groups due to DESI's target selection, hence the numbers in Summary column do not exactly equal the summation of all groups. 

Table~\ref{tab: catalogue} in Appendix~\ref{app: catalogue} provides a detailed description of our catalogue. The magnitudes in $g$, $r$ and $z$ from DESI and $W1$ and $W2$ from WISE and their corresponding errors are converted to AB magnitude system from nanomaggies given in DESI data release. We also provide the indication of group that each source belongs to.

\section{Discussions}\label{sec: discussion}
In this section, we first utilize unsupervised learning techniques to explore the feature spaces for BGS, LRG, ELG and NON sources. This analysis aims to elucidate the distinct performance across these groups and to identify potential improvements for NON sources by examining patterns within their feature space. Next, we investigate the correlation between performance and morphological classifications, explaining how different morphologies impact the accuracy of photo-$z$ estimations. Finally, we demonstrate that the precision of our photo-$z$ estimations can be further enhanced with future data releases. 
\subsection{UMAP analysis for galaxies}\label{sec: umap}

\begin{figure*}
    \centering
    \includegraphics[width=0.4\textwidth]{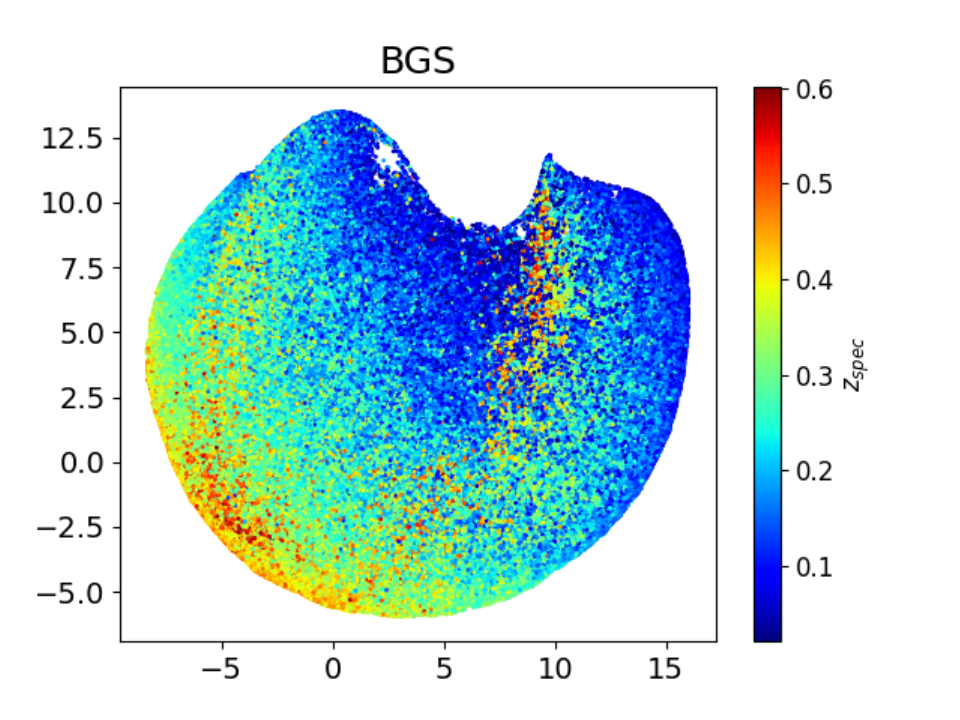}
    \includegraphics[width=0.4\textwidth]{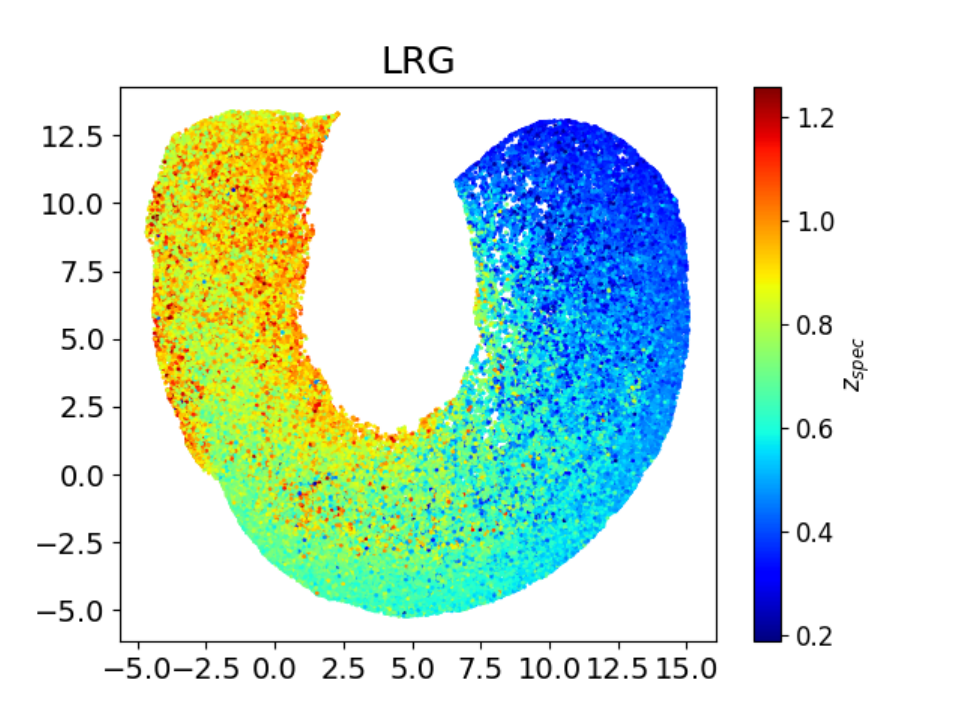}
    \\
    \includegraphics[width=0.4\textwidth]{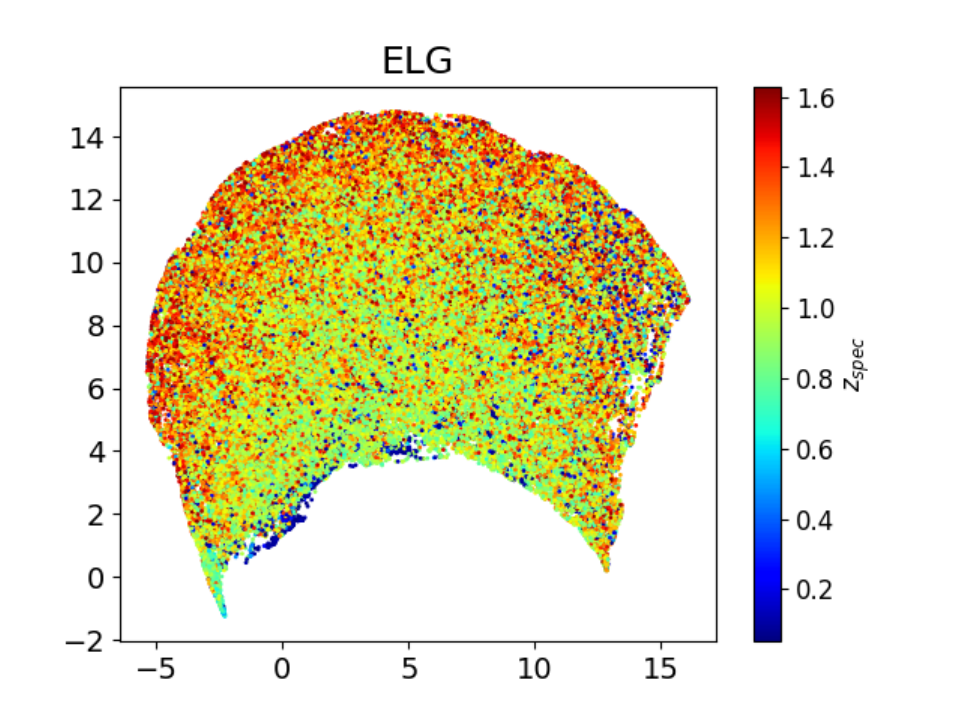}
    \includegraphics[width=0.4\textwidth]{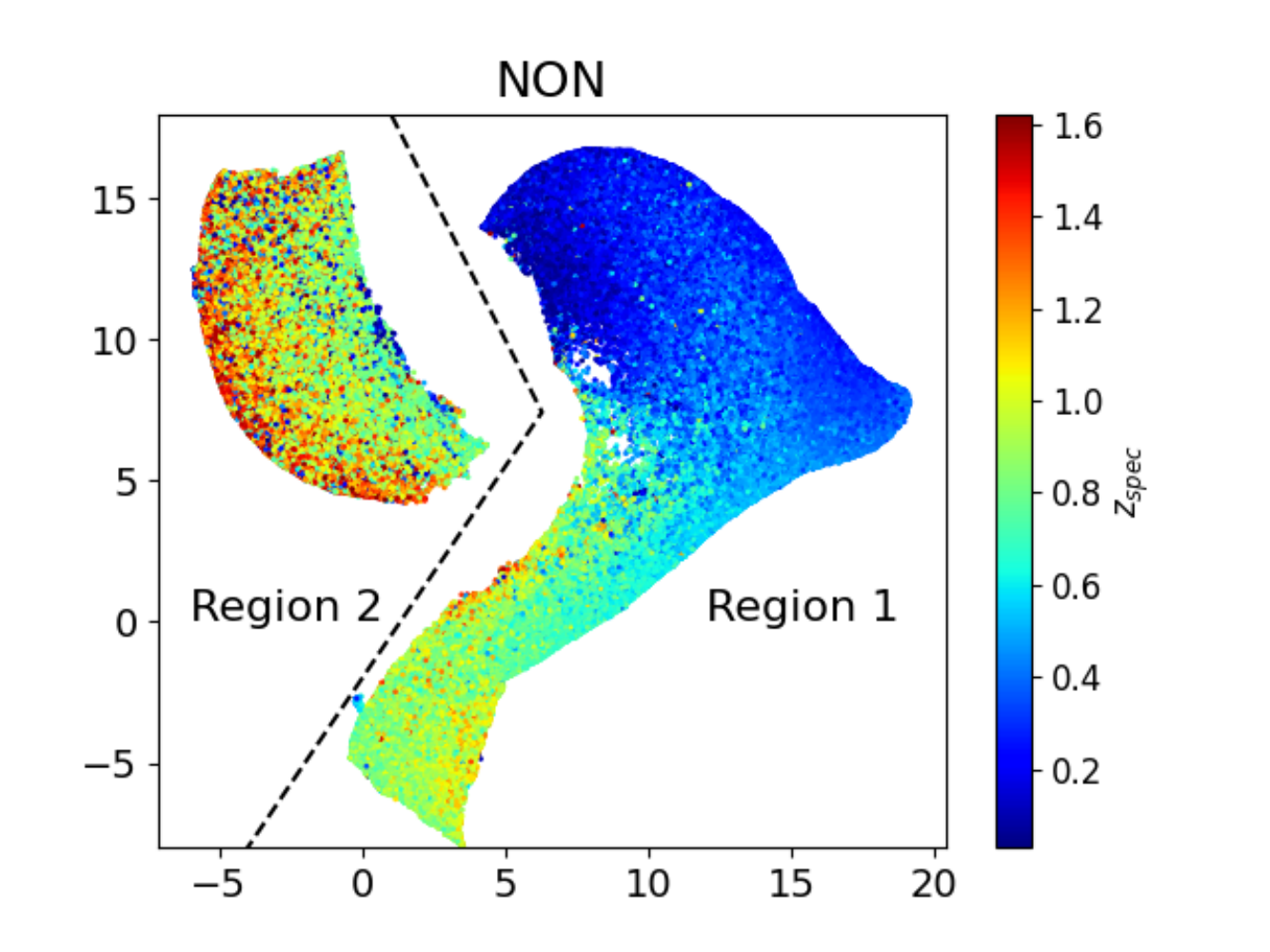}
    \caption{Two dimensional UMAP space for BGS, LRG, ELG and NON sources through magnitudes in $g, r, z W1, W2$ and half-light radius. Note that the colorbar indicates the spectroscopic redshift, and the axes are meaningless, only indicating the positions in feature space. For BGS and LRG, we can clearly recognize a correlation between redshift and positions, which explains the excellent performance for photo-$z$ estimations for these two categories. For ELG, the correlation between redshift and positions is not apparent compared to BGS and LRG, and the sources with different redshifts tend to overlap. This interprets the large bias and terrible results for ELG. The situation for NON is more complicated, with two regions disconnecting each other. {The Region 1 exhibits better correlation, while the Region 2 is much worse, showing  a similar behavior to ELG.}}
    \label{fig: umap}
\end{figure*}

In our research, we estimate photo-$z$s for galaxies directly from multi-band photometric imaging data. Our methodology aligns with other empirical approaches, which derive redshifts from photometric measurements, but uniquely incorporates morphological information extracted by convolutional neural networks. To delve deeper into the performance distinctions among the sources as discussed in Section~\ref{sec: BNN results}, we utilize Uniform Manifold Approximation and Projection (UMAP), a dimension reduction technique grounded in manifold learning and topological data analysis~\citep{2018arXiv180203426M}. UMAP can help uncover underlying patterns in data through unsupervised clustering, providing some insights on specific task. 

We employ this technique to reduce the dimension of photometric measurements in 5 bands for each source. Additionally, to mimic how our deep learning model works, we also incorporate one morphological data, the half-light radius. The two dimensional UMAP plots are illustrated in Figure~\ref{fig: umap}, where the color bar represents spectroscopic redshifts, while the axes merely indicate positions within the feature space without intrinsic meaning. 

From these plots, a clear correlation between redshift and positioning is evident for BGS and LRG, substantiating their strong photo-$z$ performance. However, for ELG, this correlation is less pronounced, with different redshifts frequently overlapping, which explains the significant bias and poor results observed for this group. We also utilize LePhare~\citep{2011ascl.soft08009A} configured with COSMOS SED sets and emission lines switched on to assess if template fitting method could yield better results for ELG. Unfortunately, this approach also fails, performing even worse than our deep learning model. As indicated in Figure~\ref{fig: elg color}, ELG exhibits a $g - r$ color around 0, suggesting a flat continuum. This characteristic likely contributes to the poor results, as both template fitting and empirical method generally rely on a gradient between different bands for accurate photo-$z$ estimations. This analysis demonstrates that the photometric redshift cannot be effectively estimated using five broad bands available from DESI and WISE. Given that DESI can accurately determine spectroscopic redshifts for ELG by resolving the [OII] doublet in their spectra, we decide not to produce the photo-$z$ catalogue for these sources. This decision reflects the inherent limitations of photo-$z$ methods for ELG, highlighting the necessity for direct spectroscopic observations to obtain reliable redshift measurement for these sources. {While for NON sources, UMAP analysis reveals a more complex structure, displaying two distinct regions within the feature space, where the Region 1 demonstrates a better correlation between feature space and redshift, whereas the Region 2 performs poorly, mirroring the challenges confronted by ELG.} This analysis interprets the reasonable performance at lower redshift and the similar bias behavior to ELG at higher redshift for NON sources as displayed in the lower right panel of Figure~\ref{fig: mnf results}. 

\begin{figure}
    \centering
    \includegraphics[width=0.5\textwidth]{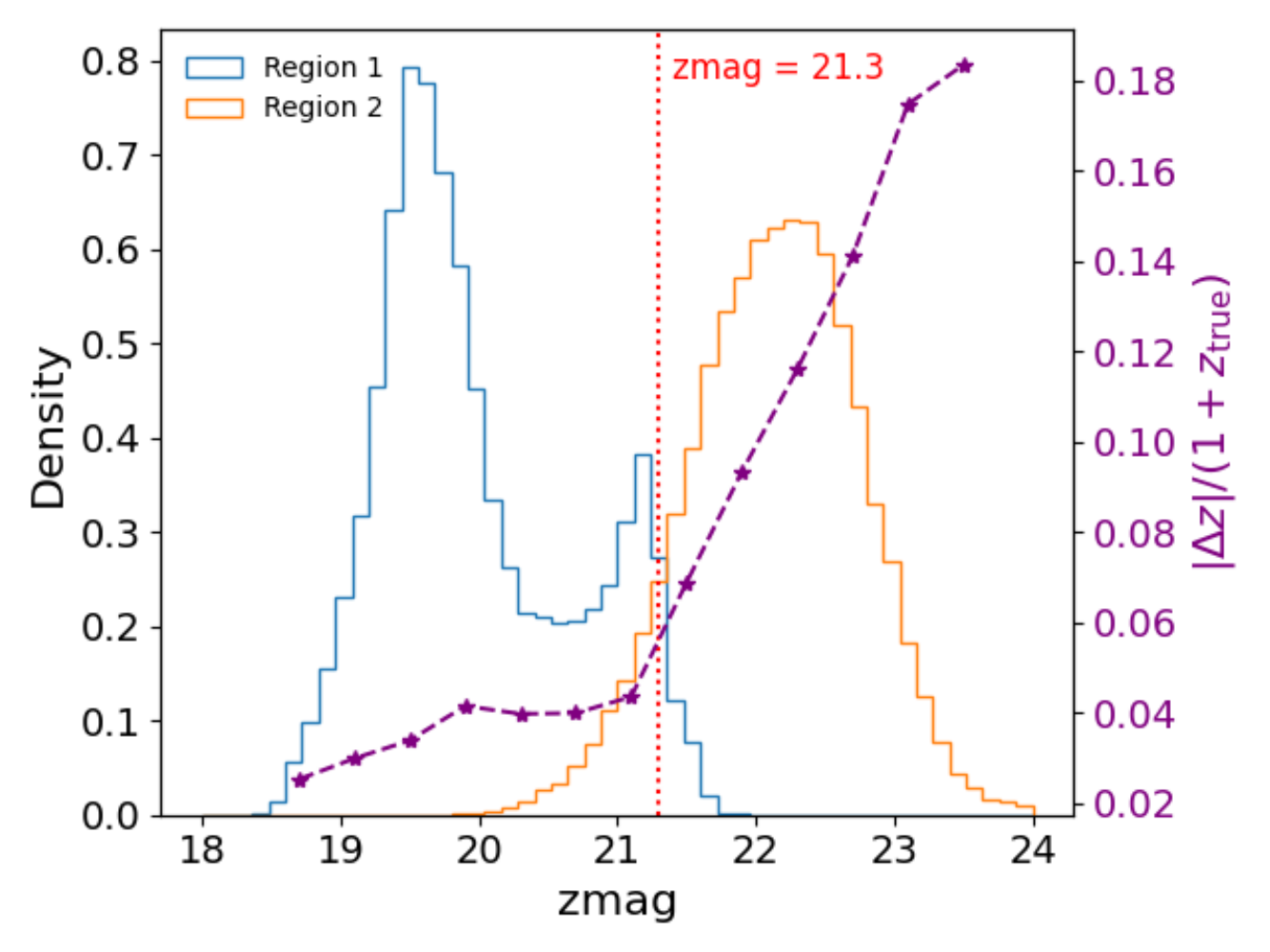}
    \caption{The $z$ band magnitude for NON sources in two regions displayed in UMAP. We recognize that the two regions can be well separated applying a magnitude cut as $z \sim 21.3$. Moreover, the absolute deviation with respect to magnitude is also display in purple color and a clear increase crossing this cut can be recognized.}
    \label{fig: non sep}
\end{figure}

Given the challenges in reliably estimating photometric redshifts for ELG, our focus shifts to the subset of NON sources that demonstrate better correlations with redshifts, as observed in the lower right panel of Figure~\ref{fig: umap}. Our analysis of the $z$ band magnitude distribution for these NON sources is detailed in Figure~\ref{fig: non sep}, which reveals that the two regions depicted in UMAP can be effectively distinguished by applying a magnitude cut at $z\sim21.3$. Additionally, this figure includes a curve in purple showing the deviation, $|\Delta z|/(1 + z_{\rm true})$, with respect to magnitude, where a noticeable increase is observed across this threshold. 

{In addition to UMAP analysis, we present the correlations between colors and spec-$z$s in Appendix~\ref{app: color-z} for BGS, LRG, ELG, and NON sources. Three colors $g-r$, $r-z$ and $W1-W2$ are considered, and NON sources are divided based on their $z$ band magnitude of 21.3. From these plots, we notice that for BGS, LRG, and NON with $z$ less than 21.3, a strong correlation exists between colors and spec-$z$. Conversely, the correlations between ELG and NON with $z$ greater than 21.3 are less pronounced, resulting in a similar conclusion to the analysis conducted using UMAP.}

\begin{figure}
    \centering
    \includegraphics[width=0.5\textwidth]{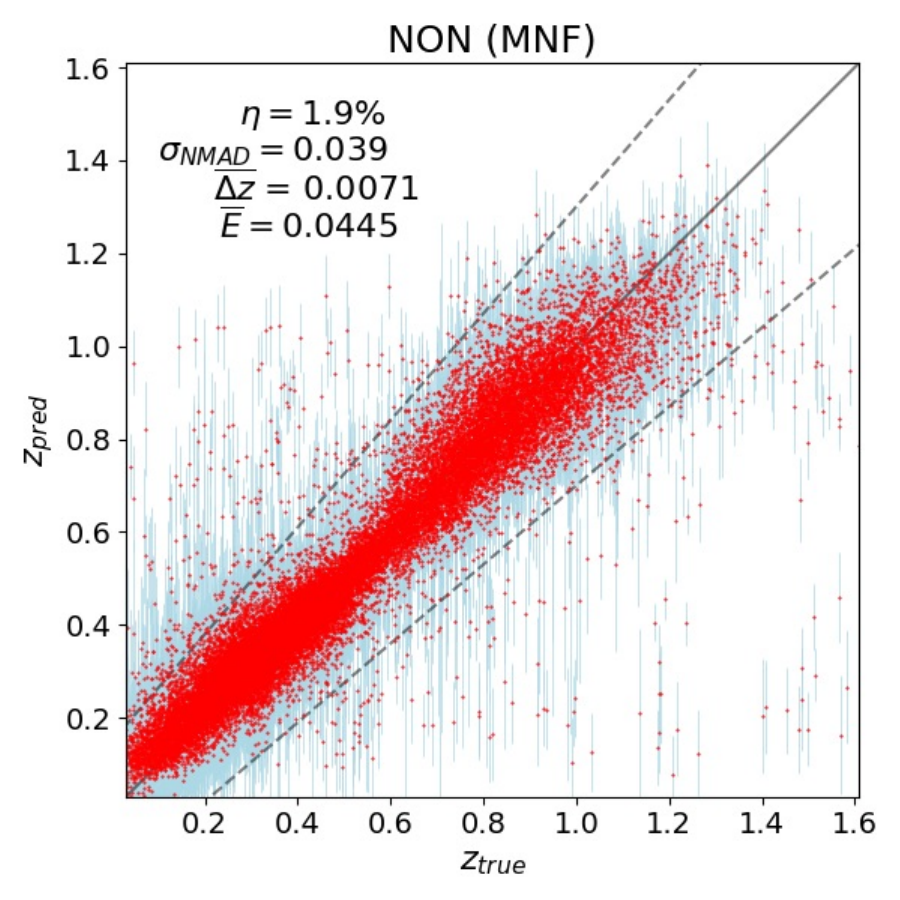}
    \caption{BNN results for NON sources with magnitude cut $z < 21.3$ applied. {We notice that the performance become significantly better but with large fraction of high redshift sources excluded compared to Figure~\ref{fig: mnf results}. }}
    \label{fig: non mnf z lim}
\end{figure}

Following this analysis, we retrain our model using NON sources with $z$ band magnitude lower than 21.3. The results, illustrated in Figure~\ref{fig: non mnf z lim}, show a significant improvement in performance, albeit at the cost of excluding a considerable number of high redshift sources. Based on these findings, our photo-$z$ catalogue includes only those NON sources with $z$ band magnitudes below 21.3, ensuring more reliable photo-$z$ estimations while acknowledging the limitations imposed by higher redshift exclusions.

\begin{table*}
\caption{{The BNN results of different morphologies for BGS, LRG, ELG and NON sources. $\eta$, $\sigma_{\rm NMAD}$, $\overline{\Delta z}$ and $\overline{E}$ indicates outlier percentage, accuracy, mean bias and mean uncertainty respectively. Additionally, $p$ shows the fraction of individual morphology for each target. }}
\label{tab: morph comp}
\begin{tabular}{|c|ccccc||ccccc|}
\hline
Morph  & \multicolumn{5}{c||}{REX}                                                                                                                                            & \multicolumn{5}{c|}{EXP}                                                                                                                                            \\ \hline
Metric & \multicolumn{1}{c|}{$\eta$}  & \multicolumn{1}{c|}{$\sigma_{\rm NMAD}$} & \multicolumn{1}{c|}{$\overline{\Delta z}$} & \multicolumn{1}{c|}{$\overline{E}$} & $p$    & \multicolumn{1}{c|}{$\eta$}  & \multicolumn{1}{c|}{$\sigma_{\rm NMAD}$} & \multicolumn{1}{c|}{$\overline{\Delta z}$} & \multicolumn{1}{c|}{$\overline{E}$} & $p$    \\ \hline
BGS    & \multicolumn{1}{c|}{0.19\%}  & \multicolumn{1}{c|}{0.022}               & \multicolumn{1}{c|}{-0.0018}               & \multicolumn{1}{c|}{0.026}          & 19.7\% & \multicolumn{1}{c|}{0.19\%}  & \multicolumn{1}{c|}{0.023}               & \multicolumn{1}{c|}{-0.0008}               & \multicolumn{1}{c|}{0.026}          & 11.6\% \\ \hline
LRG    & \multicolumn{1}{c|}{0.58\%}  & \multicolumn{1}{c|}{0.032}               & \multicolumn{1}{c|}{0.0007}                & \multicolumn{1}{c|}{0.033}          & 29.1\% & \multicolumn{1}{c|}{0.75\%}  & \multicolumn{1}{c|}{0.032}               & \multicolumn{1}{c|}{0.001}                 & \multicolumn{1}{c|}{0.036}          & 10.0\% \\ \hline
ELG    & \multicolumn{1}{c|}{16.30\%} & \multicolumn{1}{c|}{0.109}               & \multicolumn{1}{c|}{0.0176}                & \multicolumn{1}{c|}{0.117}          & 82.1\% & \multicolumn{1}{c|}{15.27\%} & \multicolumn{1}{c|}{0.100}               & \multicolumn{1}{c|}{0.0231}                & \multicolumn{1}{c|}{0.117}          & 13.2\% \\ \hline
NON    & \multicolumn{1}{c|}{1.98\%}  & \multicolumn{1}{c|}{0.042}               & \multicolumn{1}{c|}{0.0055}                & \multicolumn{1}{c|}{0.044}          & 38.8\% & \multicolumn{1}{c|}{2.57\%}  & \multicolumn{1}{c|}{0.044}               & \multicolumn{1}{c|}{0.0127}                & \multicolumn{1}{c|}{0.052}          & 21.2\% \\ \hline
Total  & \multicolumn{1}{c|}{7.39\%}  & \multicolumn{1}{c|}{0.049}               & \multicolumn{1}{c|}{0.0082}                & \multicolumn{1}{c|}{0.069}          & 36.9\% & \multicolumn{1}{c|}{3.79\%}  & \multicolumn{1}{c|}{0.037}               & \multicolumn{1}{c|}{0.0080}                & \multicolumn{1}{c|}{0.052}          & 13.5\% \\ \hline \hline
Morph  & \multicolumn{5}{c||}{DEV}                                                                                                                                            & \multicolumn{5}{c|}{SER}                                                                                                                                            \\ \hline
Metric & \multicolumn{1}{c|}{$\eta$}  & \multicolumn{1}{c|}{$\sigma_{\rm NMAD}$} & \multicolumn{1}{c|}{$\overline{\Delta z}$} & \multicolumn{1}{c|}{$\overline{E}$} & $p$    & \multicolumn{1}{c|}{$\eta$}  & \multicolumn{1}{c|}{$\sigma_{\rm NMAD}$} & \multicolumn{1}{c|}{$\overline{\Delta z}$} & \multicolumn{1}{c|}{$\overline{E}$} & $p$    \\ \hline
BGS    & \multicolumn{1}{c|}{0.14\%}  & \multicolumn{1}{c|}{0.017}               & \multicolumn{1}{c|}{-0.0014}               & \multicolumn{1}{c|}{0.019}          & 14.7\% & \multicolumn{1}{c|}{0.11\%}  & \multicolumn{1}{c|}{0.017}               & \multicolumn{1}{c|}{-0.0015}               & \multicolumn{1}{c|}{0.019}          & 54.1\% \\ \hline
LRG    & \multicolumn{1}{c|}{0.28\%}  & \multicolumn{1}{c|}{0.024}               & \multicolumn{1}{c|}{0.0039}                & \multicolumn{1}{c|}{0.027}          & 45.3\% & \multicolumn{1}{c|}{0.48\%}  & \multicolumn{1}{c|}{0.020}               & \multicolumn{1}{c|}{0.0009}                & \multicolumn{1}{c|}{0.025}          & 15.6\% \\ \hline
ELG    & \multicolumn{1}{c|}{14.18\%} & \multicolumn{1}{c|}{0.098}               & \multicolumn{1}{c|}{0.0187}                & \multicolumn{1}{c|}{0.113}          & 4.5\%  & \multicolumn{1}{c|}{19.3\%}  & \multicolumn{1}{c|}{0.100}               & \multicolumn{1}{c|}{0.0173}                & \multicolumn{1}{c|}{0.129}          & 0.2\%  \\ \hline
NON    & \multicolumn{1}{c|}{1.52\%}  & \multicolumn{1}{c|}{0.033}               & \multicolumn{1}{c|}{0.0019}                & \multicolumn{1}{c|}{0.035}          & 24.3\% & \multicolumn{1}{c|}{1.4\%}   & \multicolumn{1}{c|}{0.035}               & \multicolumn{1}{c|}{0.0116}                & \multicolumn{1}{c|}{0.050}          & 15.6\% \\ \hline
Total  & \multicolumn{1}{c|}{1.11\%}  & \multicolumn{1}{c|}{0.024}               & \multicolumn{1}{c|}{0.0024}                & \multicolumn{1}{c|}{0.030}          & 20.2\% & \multicolumn{1}{c|}{0.31\%}  & \multicolumn{1}{c|}{0.018}               & \multicolumn{1}{c|}{0.0001}                & \multicolumn{1}{c|}{0.023}          & 29.3\% \\ \hline
\end{tabular}
\end{table*}

\subsection{Correlation between photo-$z$ accuracy and morphology}\label{sec: morph class}

\begin{figure}
    \centering
    \includegraphics[width=0.5\textwidth]{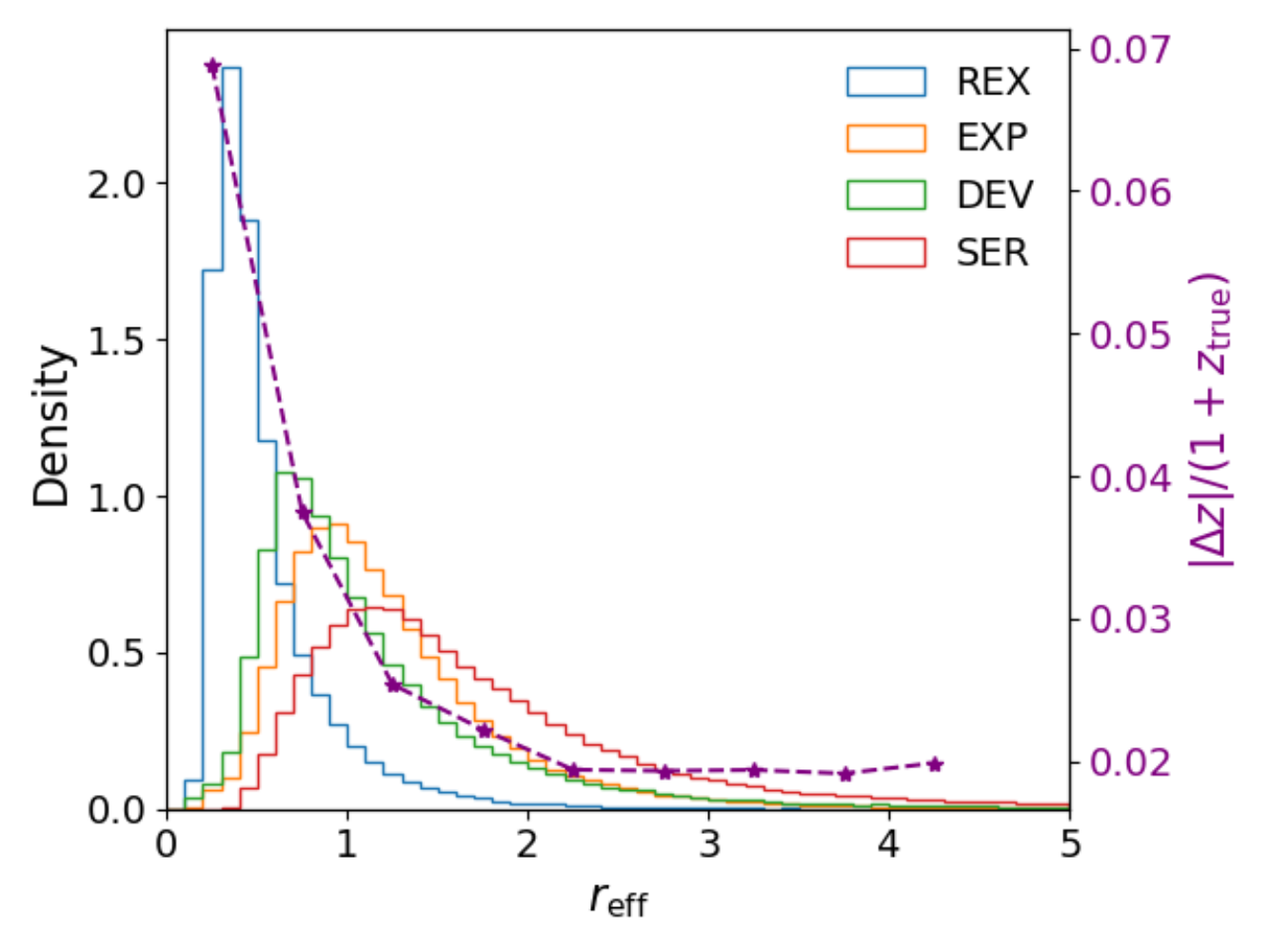}
    \caption{The distribution of half-light radius of individual morphology. Moreover, absolute deviation with respect to the radius is also displayed in purple dashed line.}
    \label{fig: r dist}
\end{figure}

DESI employs the Tracter software for morphological classification during its photometry measurements, assigning each source one of five model types: point sources (PSF), round exponential galaxies with a variable radius (REX), deVaucouleurs (DEV) profiles (elliptical galaxies), exponential (EXP) profiles (spiral galaxies), and Sersic (SER) profiles. In our analysis, we focus on galaxies classified as non-PSF, encompassing REX, DEV, EXP and SER models. 

We explore the distribution of half-light radii for these morphological types and the 
absolute deviation with respect to radius in Figure~\ref{fig: r dist}. This figure highlights a clear trend that the absolute deviation decreases with increasing radius and the SER profiles exhibit the largest radius, followed by DEV, EXP and REX. Although the radii of some galaxies have large errors, these errors do not affect the overall distribution or the trend between the deviation and radius. {Furthermore, the performance of photo-$z$ estimations of different morphological profiles for BGS, LRG, ELG and NON sources is presented in Table~\ref{tab: morph comp}, where $\eta$, $\sigma_{\rm NMAD}$, $\overline{\Delta z}$ and $\overline{E}$ indicate outlier, accuracy, mean bias and mean uncertainty respectively. Additionally, the fraction of individual morphology for each target $p$ is also shown.}

The results reveal interesting correlations between morphology and photo-$z$ performance. {Firstly, from a comprehensive perspective, the performance of the photo-$z$ and uncertainty estimations exhibits significant correlations among the four morphological types, with SER performing best followed by DEV, EXP, and REX.}
{Secondly, we find that SER profiles dominate among BGS sources, likely contributing to their superior photo-$z$ and uncertainty estimations.} SER profiles typically allow for more accurate morphological parameterization by CNNs due to their variable brightness profiles that capture galaxy structure nuances. 
{And thirdly, the prevalence of smaller-sized galaxies, DEV, among LRG could explain their diminished photo-$z$ accuracy and uncertainty estimations compared to BGS as smaller galaxies provide less structural information for CNNs to utlize effectively.} {Furthermore, the considerable majority of ELG are REX with smallest radii among these morphological types, which may hinder the extraction of detailed morphological features by CNNs, accounting for the worst photo-$z$ and uncertainties.}

{These analysis suggest a notable correlation between morphological classifications and accuracy of photo-$z$ and confidence of these estimations from galaxy images by neural networks. Galaxies characterized with more complex and larger radii as SER profiles promote the information extraction by CNN, thus yielding better result, while smaller-sized galaxies, such as those with REX profiles, tend to degrade photo-$z$ accuracy and provide poor uncertainties.}

\subsection{Potential improvement with future data release}\label{sec: more data}
\begin{figure}
    \centering
    \includegraphics[width=0.5\textwidth]{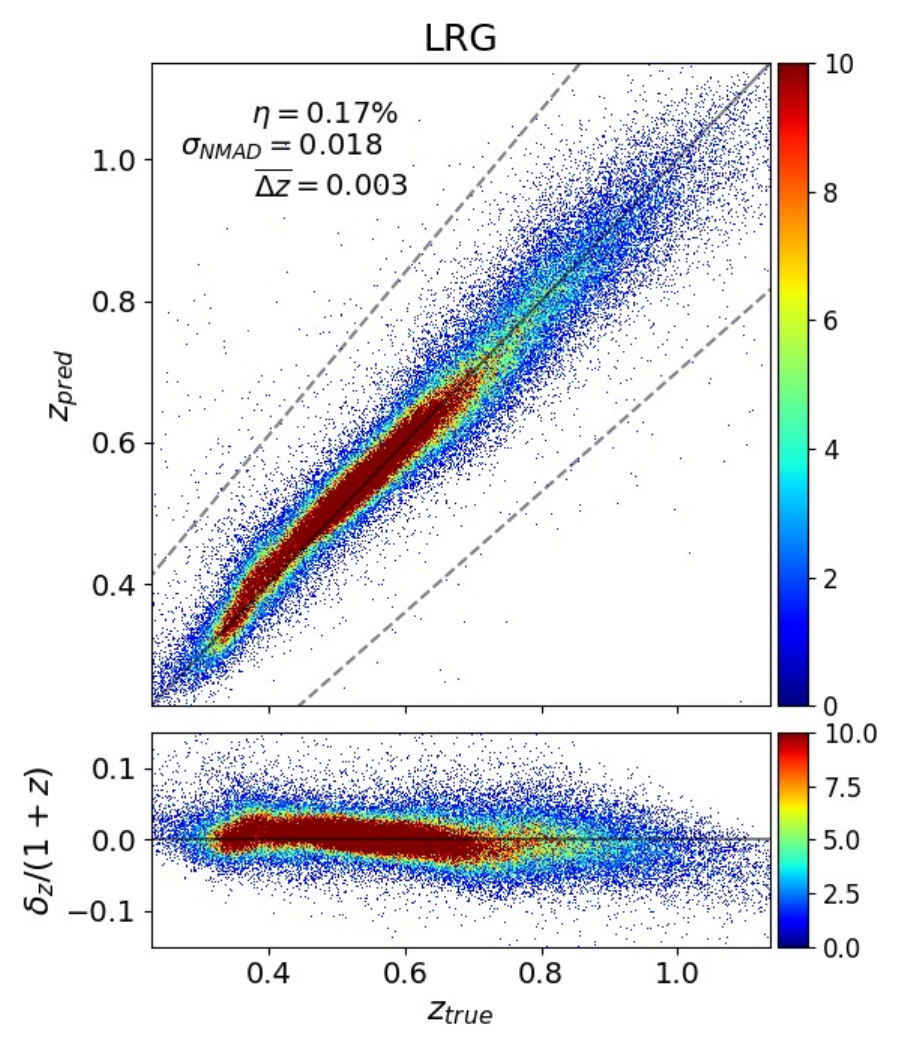}
    \caption{CNN results for LRG using $0.6$ million sources supplemented by~\citet{Zou2019}. We notice that the outlier $\eta$ and accuracy $\sigma_{\rm NMAD}$ are significantly reduced to comparable level to their work. This indicates that with future DESI data releases, our photo-$z$ estimations can definitely achieve better performance.}
    \label{fig: lrg supple}
\end{figure}

{From the comparison in Table~\ref{tab: compare}, it is evident that directly processing galaxy images via CNNs can slightly decrease the outlier fraction for photo-$z$ estimation of BGS.} Impressively, this approach requires only about one-quarter of the training data compared to methods that rely solely on photometric measurements, as reported in the work by~\citet{Zou2019}. {However, it is important to note that despite the improvement in outlier, the accuracy and mean bias achieved through our CNN remains worse and does not yet compare favorably with the results from theirs.}

As for LRG, the results significantly degenerate behind those for BGS. We attribute this result to a lack of sufficient data, which is often a critical factor in training effective machine learning models. To address this, we supplement our LRG dataset with additional sources from~\citet{Zou2019}, increasing our dataset to approximately 0.6 million sources. With this enhanced dataset, we obtain $\eta=0.17\%$ and $\sigma_{\rm NMAD}=0.018$ as shown in Figure~\ref{fig: lrg supple}. These results demonstrate that with more training data, not only is the outlier fraction substantially improved, but the accuracy is also significantly enhanced to comparable level of their work. 

While the insights gained from using an expanded dataset highlight the potential for improved photo-$z$ estimation, it is crucial to recognize that for consistency and quality consideration of spec-$z$s, our photo-$z$ catalogue remains based solely on DESI observations. The discussion here serves primarily as a proof of concept, illustrating that with future data releases from DESI, it is plausible to expect even better performance from our photo-$z$ estimations. We anticipate updating our photo-$z$ catalogue in alignment with future data releases from DESI, ensuring that our estimations leverage the most comprehensive and up-to-date observational data available. 

\section{Conclusions}\label{sec: conclusion}
This paper presents a catalogue of photometric redshifts for galaxies in DESI Legacy Surveys, encompassing totally $\sim0.18$ billion sources covering 14,000 $\deg^2$. The photometric redshifts and their uncertainties are directly estimated through galaxy images in three optical bands from DESI and two near-infrared bands from WISE employing Bayesian neural networks. The BNN is trained using the above images and high-quality spectroscopic redshifts provided in DESI Early Data Release. A key advantage of using galaxy images is the intrinsic inclusion of morphological information, which can be effectively utilized by CNNs. This approach has generally matched or even outperformed methods that rely solely on photometric measurements, particularly in reducing outlier fractions $\eta$, with fewer training samples required. 

Additionally, we find that categorizing sources into individual groups based on their characteristics and estimating their redshifts within their group can effectively produce higher precision. By adhering to DESI's main target selection criteria, we categorize the sources into four groups: BGS, LRG, ELG and NON. Our approach of separate estimations for these sources has demonstrated enhanced performance, especially for BGS and LRG sources, due to mitigation of potential confusion of sources in feature space. 

{As measured by outliers of $|\Delta z| > 0.15(1 + z_{\rm true})$, accuracy by $\sigma_{\rm NMAD}$ and mean uncertainty $\overline{E}$, we achieve low outlier percentage, high accuracy and low uncertainty: 0.14\%, 0.018 and 0.0212 for BGS and 0.45\%, 0.026 and 0.0293 for LRG respectively, and we demonstrate that an increase in training data volume will result in improved performance for both metrics.} {However, for ELG, the photo-$z$ estimations display significant scatter and bias, showing results of $>15\%$, $\sim0.1$ and $\sim0.1$ irrespective of method employed, whether SED fitting or empirical one.} As depicted by UMAP created from five magnitudes and half-light radius, the correlation between redshifts and feature space positions for ELG are less pronounced compared to BGS and LRG. And the flat continuum of ELG spectra probably aggravate the difficulty of photo-$z$ estimation. The analysis of ELG suggests that for this group of sources, spectroscopic redshifts are necessary. As for NON sources, the UMAP analysis revealed two distinct regions, with one showing strong correlation with redshifts and the other one resembling the challenges confronted with ELG. By applying a magnitude cut of $z<21.3$, these two regions can be effectively separated, and the outlier, accuracy and mean uncertainty can be significantly enhanced to 1.9\%, 0.039 and 0.0445 respectively. {Additionally, the analysis by colors vs. z plots demonstrate the same conclusion.}

Moreover, our study examined the relationship between performance and morphological classification, concluding that larger half-light radii tend to improve photo-$z$ estimation, since useful information can be better extracted by CNN. Specifically, the Sersic (SER) profile, which indicates large radii and describes more detailed structural features of galaxies, showed the best performance. Conversely, the round exponential (REX) profile, typically representing smaller-sized galaxies, performed the worst, which partially explains the inferior results for ELG, where the considerable majority are REX.

{Our findings confirms that estimating photometric redshifts directly from galaxy images using deep learning is remarkably potential by naturally incorporating morphological information.} Additionally, the source categorization based on galaxy characteristics can significantly enhance the performance of photo-$z$ estimations. Given the findings, we produce a photo-$z$ catalogue for sources in DESI Legacy Surveys. We recognize that our results are constrained by the limited dataset size available in the DESI EDR, and restricted by the five band images in Legacy Surveys. As more data become available from future DESI release and with photometric measurements from Euclid, CSST and other wide surveys involved, we plan to refine and update our photo-$z$ catalogue to further enhance its accuracy and reliability.

\section*{Acknowledgements}

XCZ and NL acknowledge the support from The science research grants from the China Manned Space Project (No. CMS-CSST-2021-A01), and The CAS Project for Young Scientists in Basic Research (No. YSBR-062), and The Ministry of Science and Technology of China (No. 2020SKA0110100). YG acknowledges the support from National Key R\&D Program of China grant Nos 2022YFF0503404 and 2020SKA0110402, and the CAS Project for Young Scientists in Basic Research (No. YSBR-092). This work is also supported by science research grants from the China Manned Space Project with grant Nos CMS-CSST-2021-B01 and CMS-CSST-2021-A01. HZ acknowledges the science research grants from the China Manned Space Project with Nos. CMS-CSST-2021-A02 and CMS-CSST-2021-A04 and the supports from National Natural Science Foundation of China (NSFC; grant Nos. 12120101003 and 12373010) and  National Key R\&D Program of China (grant Nos. 2023YFA1607800, 2022YFA1602902) and Strategic Priority Research Program of the Chinese Academy of Science (Grant Nos. XDB0550100). Z.H. acknowledges of the support from National Postdoctoral Grant Program (No GZC20232990).

\section*{Data Availability}
The 9th data release of DESI Legacy Imaging Surveys are publicly available at \url{https://www.legacysurvey.org/dr9/}~\citep{Dey2019}. The early data release of DESI is publicly available at \url{https://data.desi.lbl.gov/doc/}~\citep{2023arXiv230606308D}. The photo-$z$ catalogue produced in~\citet{Zou2019} is publicly available at \url{http://batc.bao.ac.cn/~zouhu/doku.php?id=projects:desi_photoz:}. The photo-$z$ catalogue produced in this work is publicly available at \url{https://pan.cstcloud.cn/web/share.html?hash=hUWwk1QTSjo}. All other data and code related to this publication are available upon reasonable request to the corresponding author.



\bibliographystyle{mnras}
\bibliography{example} 




\appendix

\section{MC-dropout results}\label{app: dropout}
\begin{figure*}
    \centering
    \includegraphics[width=0.4\textwidth]{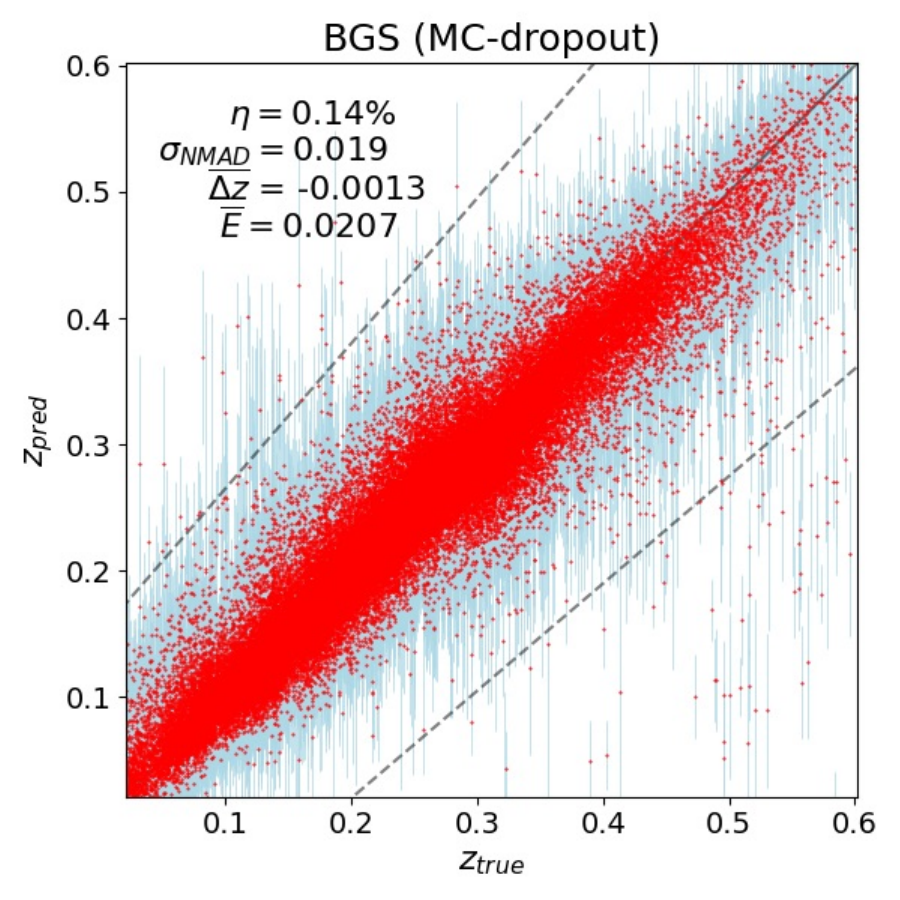}
    \includegraphics[width=0.4\textwidth]{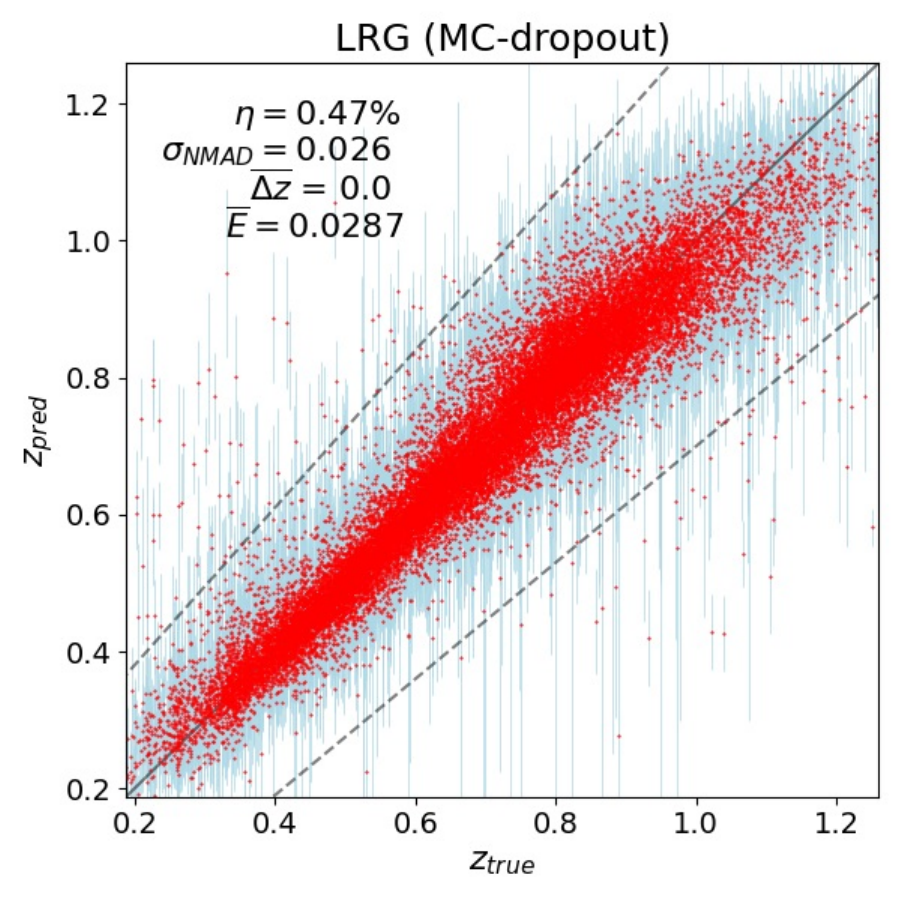}\\
    \includegraphics[width=0.4\textwidth]{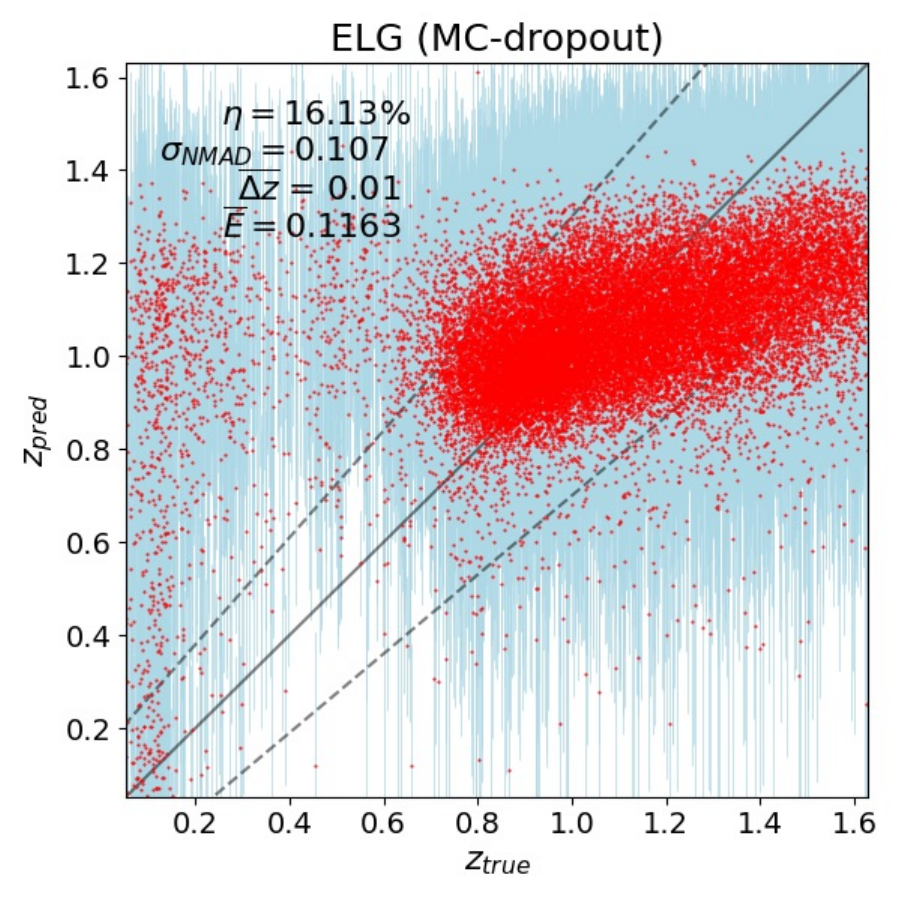}
    \includegraphics[width=0.4\textwidth]{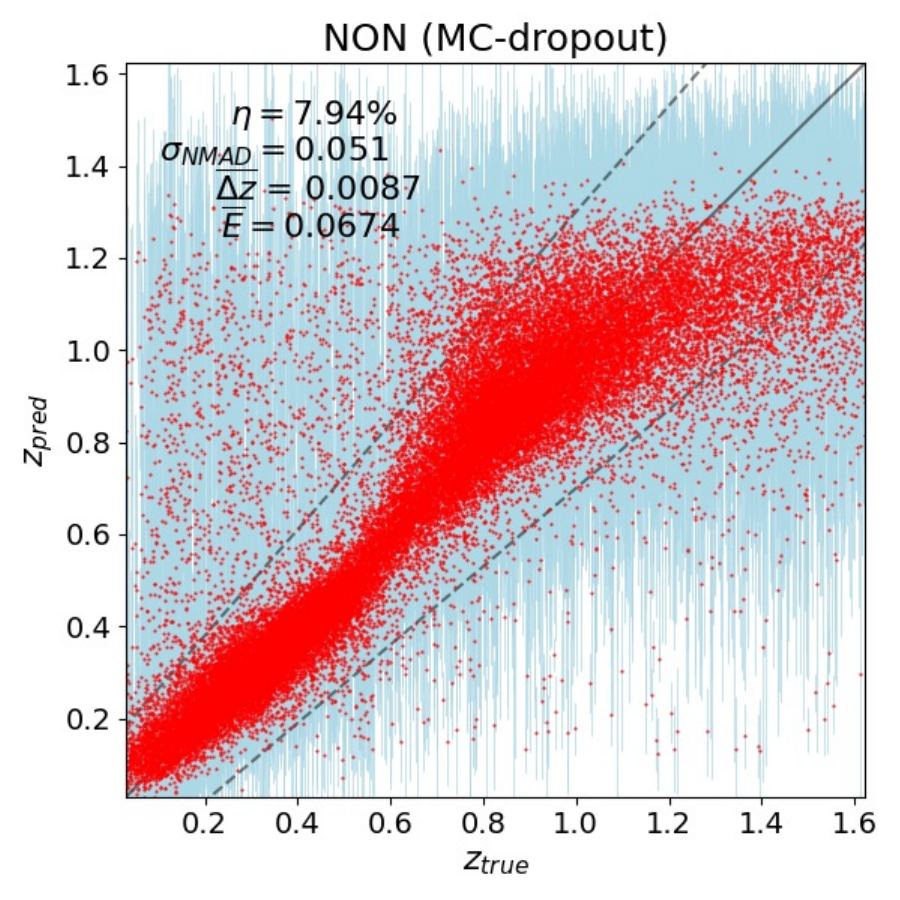}
    \caption{The results of BNN implemented by MC-dropout for BGS, LRG, ELG and NON targets. $\eta$, $\sigma_{\rm NMAD}$, $\overline{\Delta z}$ and $\overline{E}$ indicates outlier percentage, accuracy, mean bias and mean uncertainty respectively. }
    \label{fig: dropout results}
\end{figure*}

The results of BNN implemented by MC-dropout are displayed in Figure~\ref{fig: dropout results}. We notice that the performance for these sources are similar to that by MNF shown in Figure~\ref{fig: mnf results}. Considering the calibration results provided in Figure~\ref{fig: reliability diagram}, BNN using MNF layers is recommended and utilized to produce our photo-$z$ catalogue. 

\section{Colors vs. specz}\label{app: color-z}
The correlations between colors and spec-$z$s are displayed in Figure~\ref{app: color-z}. From these plots, we can obtain same conclusions to the analysis of UMAP discussed in Section~\ref{sec: umap}. 
\begin{figure*}
    \centering
    \includegraphics[width=0.75\linewidth]{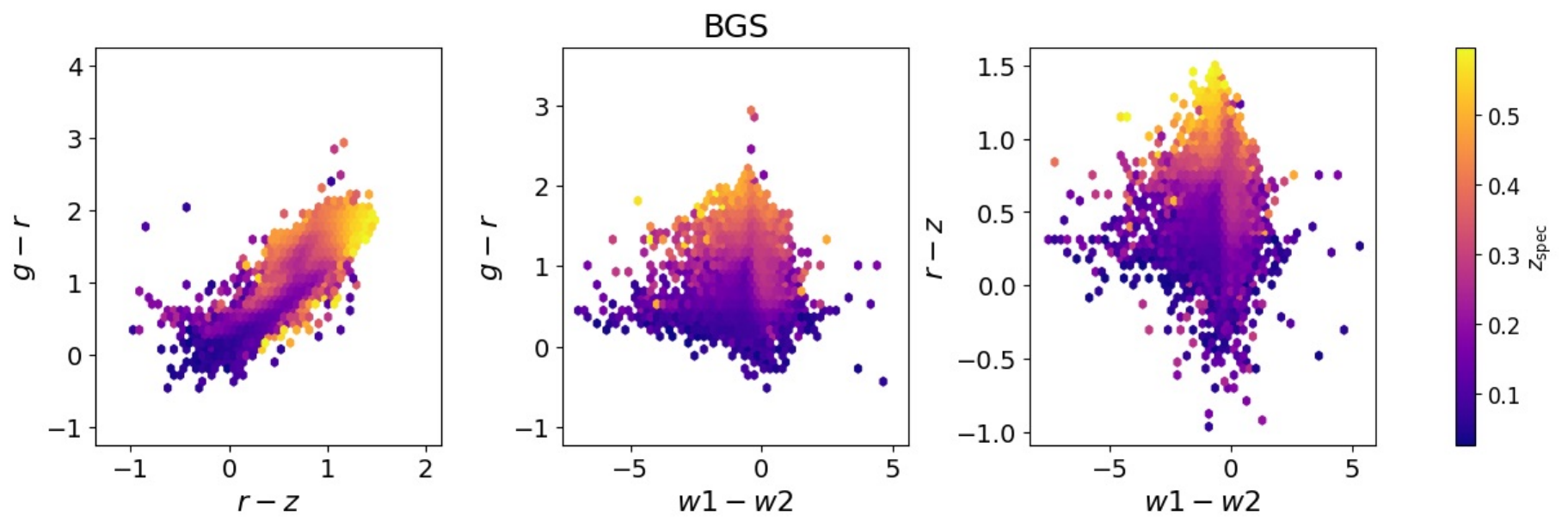}
    \includegraphics[width=0.75\linewidth]{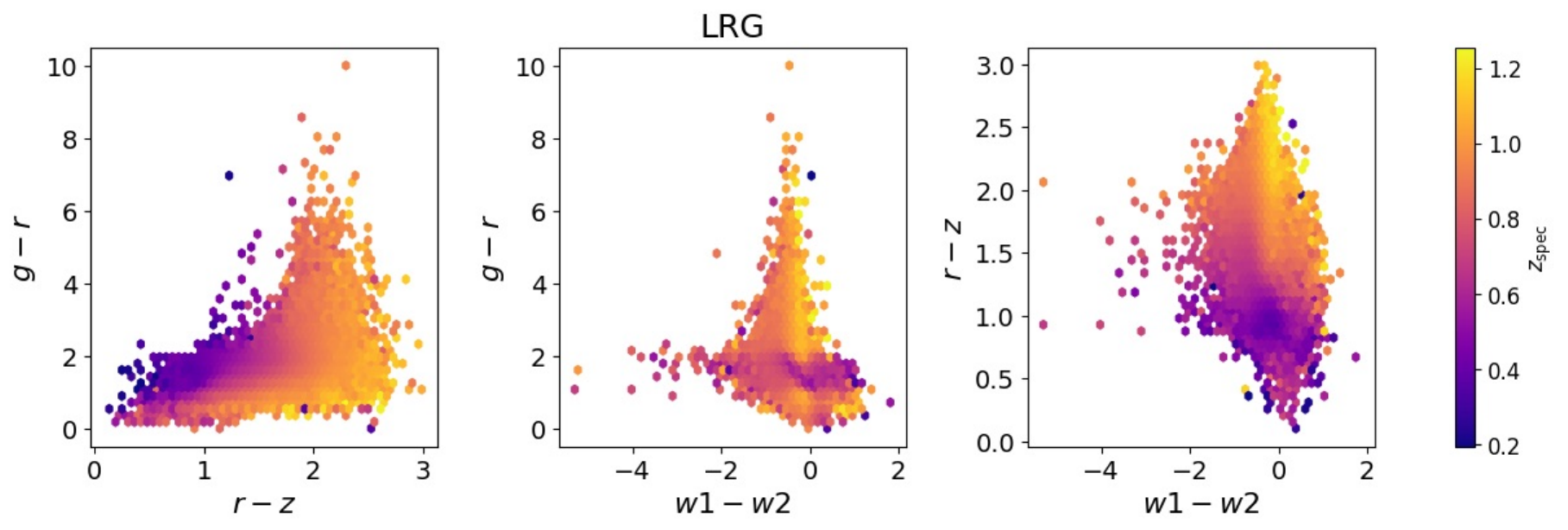}
    \includegraphics[width=0.75\linewidth]{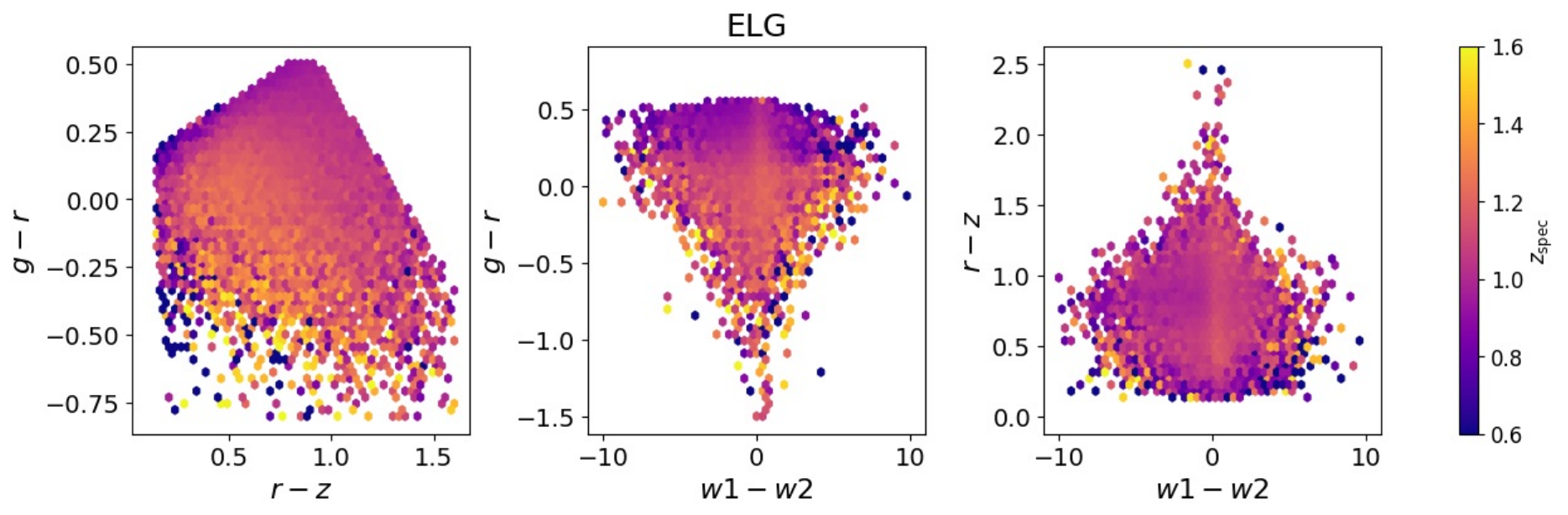}
    \includegraphics[width=0.75\linewidth]{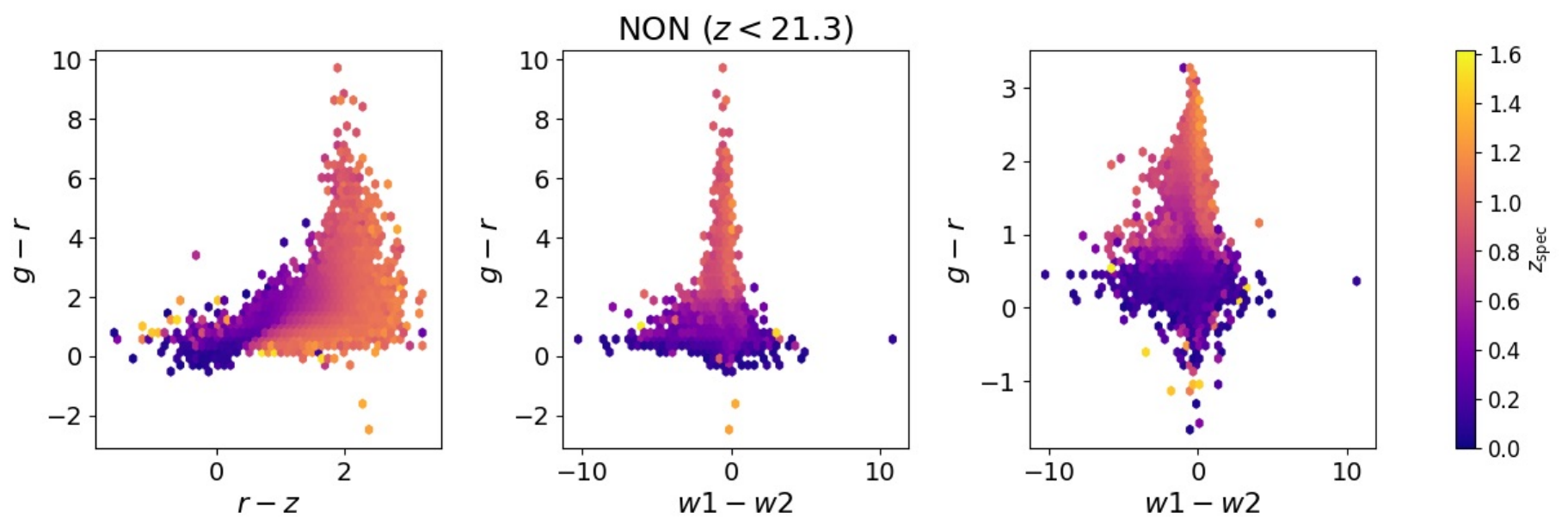}
    \includegraphics[width=0.75\linewidth]{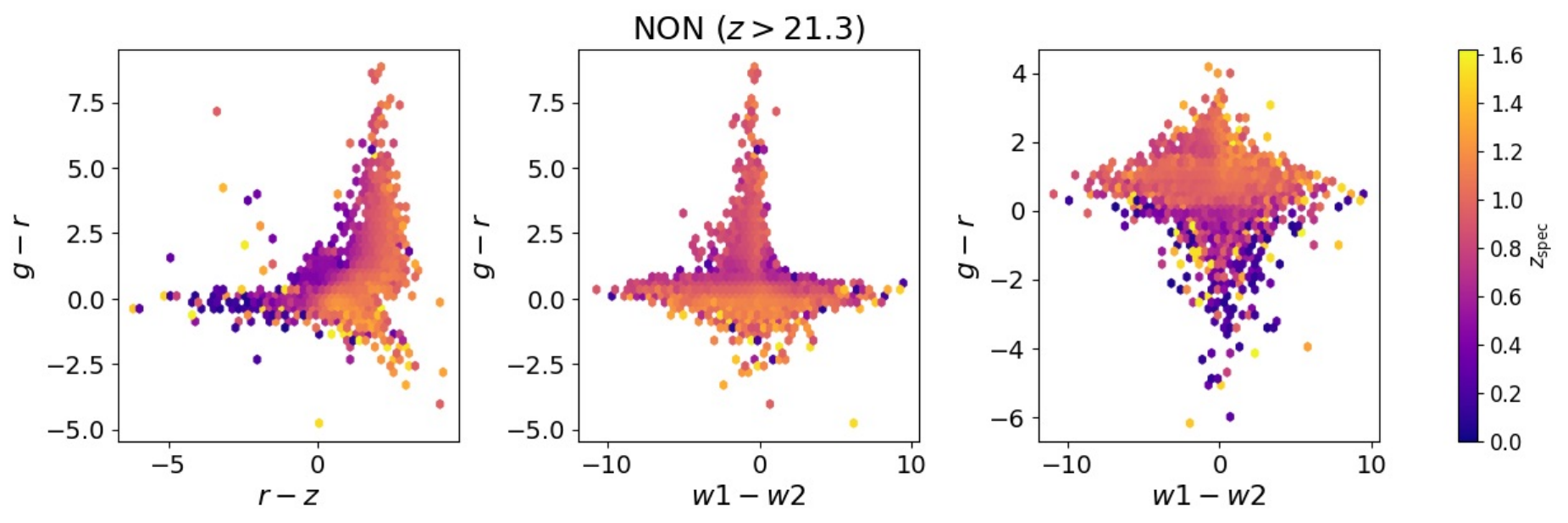}
    \caption{Correlations between colors and spec-$z$ for BGS, LRG, ELG and NON sources. NON sources are divided into two parts considering the $z$ band magnitude threshold. The color bar indicates the spec-$z$.}
    \label{fig: color_z}
\end{figure*}

\section{Description of our photo-$z$ catalogue}\label{app: catalogue}

\newcolumntype{b}{X}
\newcolumntype{s}{>{\hsize=.5\hsize}X}

\begin{table*}
\caption{Description of our photo-$z$ catalogue.}
\label{tab: catalogue}
\begin{tabularx}{\textwidth}{sssb}
  \textbf{Column} & \textbf{Type} & \textbf{Unit} & \textbf{Description} \\
\hline\hline
RA          & float64 & degree & Right ascension in J2000                 \\ 
Dec         & float64 & degree & Declination in J2000                     \\ 
type        & str & ... & Morphological type: REX, EXP, DEV or SER \\ 
shape\_r    & float64 & arcsec & Half-light radius                         \\ 
MAG\_G      & float64 & mag & g-band magnitude                         \\ 
MAG\_R      & float64 & mag & r-band magnitude                         \\ 
MAG\_Z      & float64 & mag & z-band magnitude                         \\ 
MAG\_W1     & float64 & mag & W1-band magnitude                        \\ 
MAG\_W2     & float64 & mag & W2-band magnitude                        \\ 
MAGERR\_G   & float64 & mag & g-band magnitude error                   \\ 
MAGERR\_R   & float64 & mag & r-band magnitude error                   \\ 
MAGERR\_Z   & float64 & mag & z-band magnitude error                   \\ 
MAGERR\_W1  & float64 & mag & W1-band magnitude error                  \\ 
MAGERR\_W2  & float64 & mag & W2-band magnitude error                  \\ 
FIBERMAG\_G & float64 & mag & g-band fiber magnitude                   \\ 
FIBERMAG\_R & float64 & mag & r-band fiber magnitude                   \\ 
FIBERMAG\_Z & float64 & mag & z-band fiber magnitude                   \\ 
BGS         & bool    & ... & If BGS sample                            \\ 
LRG         & bool    & ... & If LRG sample                            \\ 
NON         & bool    & ... & If NON sample                            \\ 
zphot       & float64 & ... & Estimated photometric redshift           \\ 
zerr        & float64 & ... & Estimated photometric redshift error     \\ 
zspec       & float64 & ... & DESI spectroscopic redshift if available \\ \hline
\end{tabularx}
\end{table*}


\bsp	
\label{lastpage}
\end{document}